\documentclass[lnicst]{svmultln}
\usepackage{graphicx}
\usepackage{makeidx}
\begin{document}
\mainmatter
\title{VeReMi: A Dataset for Comparable Evaluation of Misbehavior Detection in VANETs}
\titlerunning{VeReMi: Dataset for MDS evaluation in VANETs}
\author{Rens W. van der Heijden \and Thomas Lukaseder \and Frank Kargl}
\authorrunning{R. W. van der Heijden et al.}
\institute{Ulm University, Albert-Einstein-Allee 11, 89081, Ulm, Germany,\\
\email{rens.vanderheijden@uni-ulm.de}, \email{thomas.lukaseder@uni-ulm.de}, \email{frank.kargl@uni-ulm.de} }

\maketitle
\begin{abstract}
Vehicular networks are networks of communicating vehicles, a major enabling technology for future cooperative and autonomous driving technologies.
The most important messages in these networks are broadcast-authenticated periodic one-hop beacons, used for safety and traffic efficiency applications such as collision avoidance and traffic jam detection.
However, broadcast authenticity is not sufficient to guarantee message correctness.
The goal of misbehavior detection is to analyze application data and knowledge about physical processes in these cyber-physical systems to detect incorrect messages, enabling local revocation of vehicles transmitting malicious messages.
Comparative studies between detection mechanisms are rare due to the lack of a reference dataset.
We take the first steps to address this challenge by introducing the Vehicular Reference Misbehavior Dataset (VeReMi) and a discussion of valid metrics for such an assessment.
VeReMi is the first public extensible dataset, allowing anyone to reproduce the generation process, as well as contribute attacks and use the data to compare new detection mechanisms against existing ones.
The result of our analysis shows that the acceptance range threshold and the simple speed check are complementary mechanisms that detect different attacks.
This supports the intuitive notion that fusion can lead to better results with data, and we suggest that future work should focus on effective fusion with VeReMi as an evaluation baseline.
\end{abstract}

\section{Introduction}
Vehicular Ad-hoc Networks (VANETs) have received extensive attention in the research community in the past two decades as a potential enabling technology for improved road safety and efficiency.
These networks, consisting of vehicles with ad-hoc wireless communication modules, are gaining importance in the context of cooperative autonomous driving applications.
The idea is that communication can significantly improve autonomous driving by essentially increasing the availability of information within the vehicle.
However, for these applications to work correctly, this information needs to be authenticated and verified for correctness~\cite{CSUR-meta-vanet}.
Standardization agencies have already defined cryptographic (IEEE 1609.2), communication (IEEE 802.11p, IEEE 1609), and application (ITS-G5, SAE J2735) standards, but addressing the correctness of the transmitted data has largely been a research issue.
Cryptographic solutions (e.g., vehicular PKIs) only provide message integrity, and do not ensure message correctness; detecting the lack of correctness in authentic messages is referred to as \emph{misbehavior detection}.
These are typically classified into \emph{data-centric} and \emph{node-centric}~\cite{vanderHeijden-survey}, depending on the semantics of the decision: in data-centric detection, the data is \emph{reliable}, while in node-centric detection, the sender is \emph{trustworthy} (and thus the messages sent by it should be trusted).

There are many remaining research challenges in the area of misbehavior detection for VANETs.
For example, similar to the area of intrusion detection, it is intuitively obvious that it is hard to build a single detection mechanism that detects all possible attacks.
Instead, many proposals aim to either detect specific attacks~\cite{vanderHeijden2016-vtcfall} (i.e., particular types of behavior that are malicious), or they try to protect a specific application by structuring the checks such that only correct messages are accepted~\cite{Stuebing-Kalman}.
Many authors have proposed to apply some type of data fusion as a tool to combine information from multiple sources~\cite{Dietzel2014-smartvehicles,Raya-INFOCOM-2008,vanderHeijden2016-vtcfall}, but it is not well-studied how individual detection mechanisms compare.
In this article, we inform this discussion with data, and argue that it is necessary to have a clear understanding of how mechanisms behave to maximize fusion performance.
For this purpose, we also introduce a new metric that can be used to study the weaknesses and strengths of specific mechanisms by looking at how their error rates are distributed over the detecting vehicles.
If the errors are concentrated in a certain area, one can either redesign the mechanism or use a situation detection mechanism (as suggested by our previous work).
In this entire process, the essential step is a large common dataset that can be shared as input for multiple mechanisms; this dataset is the VeReMi dataset (Vehicular Reference Misbehavior Dataset), one of the two main contribution of this work.
To the best of our knowledge, this is the first of such datasets that is publicly available.

To illustrate the need for such a public dataset, it is worth looking at the different approaches taken to evaluate VANET applications, which can be categorized in three groups: real-world field studies, analytical models and simulations.
Field studies are effective for some scenarios, but especially for security and for large scale applications, this leads to prohibitive cost, especially for attacks aiming to disrupt traffic and cause accidents.
Analytical models typically assume significant simplifications to keep the evaluation manageable.
Therefore, simulation studies are often used as the primary evaluation methods for VANETs~\cite{CSUR-meta-vanet}.
Even when using simulation, the computational cost for a representative analysis is significant, suggesting that common datasets could be useful.
The state of datasets for intrusion detection evaluation is best illustrated by a recent survey by Mitchell~\&~Chen~\cite{CSUR-CPS}, who analysed intrusion detection techniques for cyber-physical systems, which is closely related to VANETs.
Out of 30 ideas discussed in~\cite{CSUR-CPS}, 4 used a public dataset, 22 papers did not release their dataset, and 4 did not use any dataset at all.
For VANETs specifically, even releasing source code is uncommon, and sometimes it is not even declared which tools are used for simulation~\cite{Joerer-simulation-survey}.
Although some authors in this field now have started to release material for reproducibility~\cite{Kumar2017-wisec,vanderHeijden2017-vnc}, this is still highly uncommon~\cite{jimenez-popper}.
This is also a challenge for the security community within VANET research; this work requires both reliable, representative VANET simulations, and public attacker implementations, to enable comparisons between different detectors.
This paper aims to address this need and push the community towards a more rigorous, scientifically valid approach to meet these and future challenges within our field.

\begin{figure}
  \centering
  \includegraphics[width=1\columnwidth]{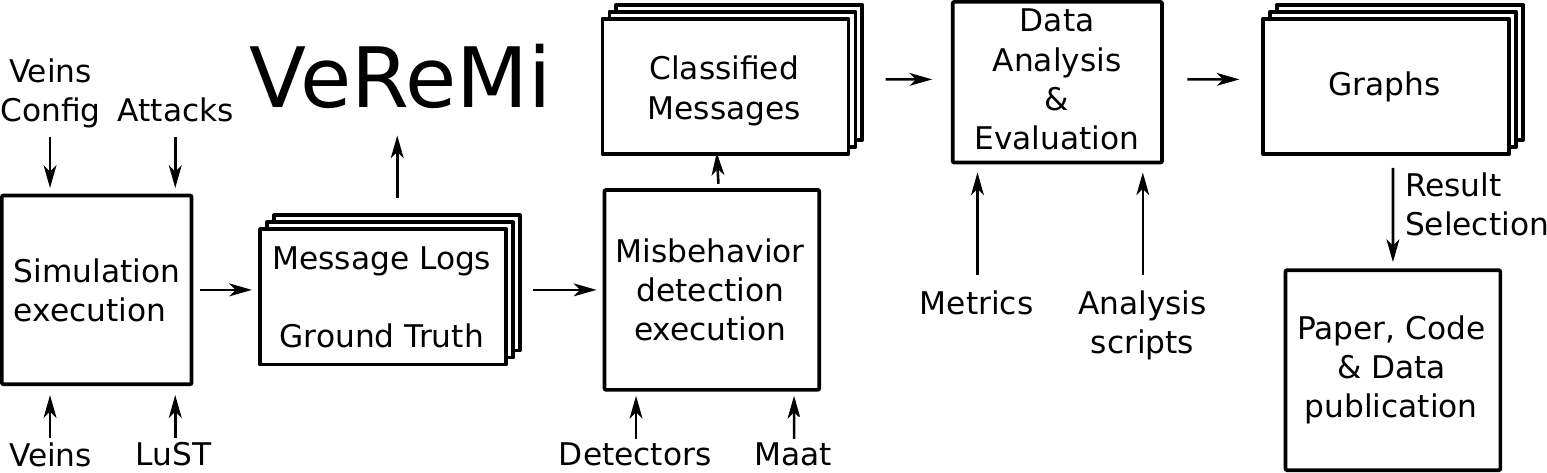}
  \caption{Evaluation workflow}\label{fig:workflow}
\end{figure}

There are many different methodologies that can be used to evaluate misbehavior detection systems (and indeed, intrusion detection in general).
In this paper, we focus on the evaluation workflow shown in Figure~\ref{fig:workflow}, which basically consists of a system simulation step, a detection step, and an analysis step.
In the system simulation step, a scenario (with or without attack(s)) is executed, and message reception is logged; in the detection step, a detection system is fed with the corresponding message logs, and the analysis step consists of computing relevant metrics and visualizing the results.
Our simulations are performed within the LuST scenario~\cite{Codeca-LuST}, using VEINS~\cite{Sommer-VEINS} for the simulation of vehicles; more details are provided in Section \ref{sec:dataset}.
The message classification process is done using our evaluation framework Maat, which executes multiple parallel detectors, as discussed in detail in Section \ref{sec:detection}.
This workflow is particularly effective for the evaluation of a broad spectrum of scenarios, and can be used to estimate the overall detection performance in a potential system deployment.
For some evaluation goals (e.g., intrusion response effectiveness), where detection must be part of the system execution, the simulation and detection steps should be combined.
In this paper, we focus on study designs that can be performed independently of the system simulation.

In summary, this paper has two major contributions; in Section~\ref{sec:dataset}, we introduce a dataset that can serve as a broad baseline for misbehavior detection evaluations, while Section~\ref{sec:metrics} describes how to aggregate and present the results.
We then show how to apply this dataset in our second major contribution, which is a broad evaluation of plausibility mechanisms proposed by previous authors in Section~\ref{sec:detection}.
We conclude with a discussion of future work in Section~\ref{sec:conclusion}.

\section{Dataset}
\label{sec:dataset}

The first contribution of this paper is a dataset intended to provide a common base line for misbehavior detection evaluation.
Previous studies have always relied on individually designed simulation studies: although this has the advantage of customizable attacks and flexibility with respect to the specifics of the scenario, it makes it difficult to compare mechanisms with each other.
The purpose of our dataset is to provide an initial baseline with which detection mechanisms can be compared.
This reduces the time required for researchers to perform high-quality simulation studies, and it makes it easier for readers to compare the results of different papers.
We acknowledge that no dataset can completely replace a detailed analysis of a detection mechanism; however, the current state of the art, where a comparison with any other scheme requires a time-intensive and error-prone re-implementation of every scheme, is unacceptable.
Our dataset will provide the first step towards a comprehensive evaluation methodology for this field.

The dataset we introduce essentially consists of message logs for every vehicle in the simulation and a ground truth file that specifies the attacker's behavior.
Local information from the vehicle is included through messages of a different type (representing periodic messages from a GPS module in the vehicle).
Any detector can thus read the sequence of messages and determine the reliability of every message (or a subset thereof).
Our dataset and the source code used to generate it \footnote{\url{https://veremi-dataset.github.io/}} is publicly available, and consists of 225 individual simulations, with 5 different attackers, 3 different attacker densities, 3 different traffic densities, and 5 repetitions for each parameter set (with different random seeds).
A detailed discussion of these aspects and the choices made in the generation process is provided below.
Note that anyone can reproduce and extend our dataset in a consistent way using the provided source code, enabling anyone to extend the evaluation of any detector that was studied with VeReMi.

\subsection{Scenario Selection}

The purpose of our dataset is to provide a holistic basis for evaluation of misbehavior detection mechanisms, rather than a specific traffic situation that works well or poorly for a specific mechanism.
This is aimed to reduce unintentional selection bias based on properties of the mechanism and the scenario, sacrificing the level of detail with which individual scenarios are studied.
The alternative approach that is often taken is to pick a few specific traffic scenarios to be studied (e.g., congested highways, free-flowing traffic in a Manhattan grid setting) and analyze these in detail.
This provides detailed information on mechanism behavior, but relies on a lot of manual decision making, making fair comparisons between mechanisms difficult.
We instead focus on how mechanisms behave in a variety of different scenarios.
To this end, we provide a much larger dataset that can be used to assess the overall performance, before looking at individual scenarios to provide specific improvements for specific detection mechanisms.
In order to achieve this, we selected a representative sample of the entire simulation scenario, based on the included road types and the associated traffic densities.

Our work is based on the Luxembourg traffic scenario (LuST), originally introduced by Codeca~et~al.~\cite{Codeca-LuST}, who aimed to provide a comprehensive scenario for evaluation of VANET applications.
Although this scenario is very suitable for traffic engineering, the simulation cost for the simulation of a city-scale VANET over multiple hours is prohibitive\footnote{For illustration purposes; our 100 second excerpt of the scenario at high densities contains hundreds of vehicles and runs for a few hours -- a significant part of this cost is the realistic simulation of signals bouncing off the ground and various buildings.}.
For this reason, reproduction of an entire study performed by other research groups is quite rare in our community -- most papers that reference results from other articles are follow-up work.
This is where our dataset comes in: it provides a simple message stream per vehicle, making it much easier to reproduce detection studies.
Table \ref{tab:simulation} describes some core parameters of the simulation -- more information can be obtained in the OMNeT++ configuration file in our source code.

\begin{table}
  \centering
  \caption{Simulation Parameters}\label{tab:simulation}
  \begin{tabular}{l | l | l}
    Parameter & Value & Notes \\ \hline
    Mobility & SUMO LuST (DUA static) & \cite{Codeca-LuST}\\
    Simulation start & (3,5,7)h & controls density \\
    Simulation duration & 100s & \\
    Attacker probability & (0.1, 0.2, 0.3) & attacker with this probability \\
    Simulation Area & 2300,5400-6300,6300 & various road types \\
    Signal interference model & Two-Ray Interference & VEINS default\\
    Obstacle Shadowing & Simple & VEINS default\\
    Fading & Jakes & VEINS default\\
    Shadowing & Log-Normal & VEINS default\\
    MAC implementation & 802.11p & VEINS default\\
    Thermal Noise & -110dbm & VEINS default\\
    Transmit Power & 20mW & VEINS default\\
    Bit rate & 6Mbps & VEINS default (best reception)\\
    Sensitivity & -89dBm & VEINS default\\
    Antenna Model & Monopole on roof & VEINS default\\
    Beaconing Rate & 1Hz & VEINS default\\
  \end{tabular}
\end{table}

\subsection{Attacks \& Implementation}

We implemented an initial set of attacks associated with position falsification, the type of attack that is most well-studied in our field (and for which many mechanisms have been designed~\cite{vanderHeijden-survey}).
Rather than implement a broad set of attacks, we focused on this specific attack to show the efficacy of our approach.
We foresee that other researchers can contribute new attack implementations and corresponding datasets to the central VeReMi repository, which we will maintain.
By focusing on a specific attack in this paper, we show how VeReMi is useful for other researchers and provide an initial starting point for the community.
Since the data is published as a list of message logs, which include speed, claimed transmission time, reception time, position, and RSSI for each receiver, it is easy to take a newer version of VeReMi and run it through detectors that have already been published.
This enables researchers to directly compare their detector against existing ones, and any new attack against a variety of detectors (as long as their source code is published).

The attackers we implement are the constant attacker, the constant offset attacker, the random attacker, the random offset attacker, and the eventual stop attacker.
The constant attacker transmits a fixed, pre-configured position; the constant offset attacker transmits a fixed, pre-configured offset added to their actual position; the random attacker sends a random position from the simulation area; the random offset attacker sends a random position in a preconfigured rectangle around the vehicle; the eventual stop attacker behaves normally for some time, and then attacks by transmitting the current position repeatedly (i.e., as if it had stopped).
The random attacks (4 and 8) take a new random sample for every message.
The parameters for our attacks are shown in Table \ref{tab:attacks}; the numbers are based on previous work~\cite{vanderHeijden2016-vtcfall}.

\begin{table}
  \centering
  \caption{Attacker parameters}\label{tab:attacks}

  \begin{tabular}{l | l | l}
    ID & Attack & Parameters\\
    \hline
    1 & Constant & $x=5560, y=5820$\\
    2 & Constant offset &$\Delta x = 250, \Delta y = -150$\\
    4 & Random & uniformly random in playground \\
    8 & Random offset& $\Delta x,\Delta y$ uniformly random from $ [-300,300] $\\
    16& Eventual stop& stop probability $+=0.025$ each position update ($10Hz$) \\
  \end{tabular}
\end{table}

\subsection{Characteristics}

The dataset consists of a total of 225 simulation executions, split into three density categories.
The low density (corresponding to a run starting at 3:00) has 35 to 39 vehicles, while the medium density (a run at 5:00) has between 97 and 108 vehicles, and the high density (7:00) has between 491 and 519 vehicles.
Out of these vehicles, a subset is malicious: this decision is made by sampling a uniform distribution ($[0,1]$) and comparing it to the attacker fraction parameter, essentially assigning each vehicle to be an attacker with that probability.
All of the vehicles classified as attacker execute the same attack algorithm (described in the previous section).
Each receiver generates a reception log containing all periodic position updates generated by SUMO ($10Hz$) and all received messages (i.e., beacons from  other vehicles).
Each of these log entries contains a reception time stamp, the claimed transmission time, the claimed sender, a simulation-wide unique message ID, a position vector, a speed vector, the RSSI, a position noise vector and a speed noise vector.
In addition, a ground truth file is updated whenever a message is sent by any vehicle: this file contains the transmission time, sender, attacker type, message ID, and actual position/speed vectors.
The attacker type is set to 0 for legitimate vehicles.
The following describes the dimensions of the VeReMi dataset in terms of messages and reception events per density.

\begin{figure}
\centering
  \parbox{0.495\textwidth}{
    \includegraphics[width=0.5\textwidth]{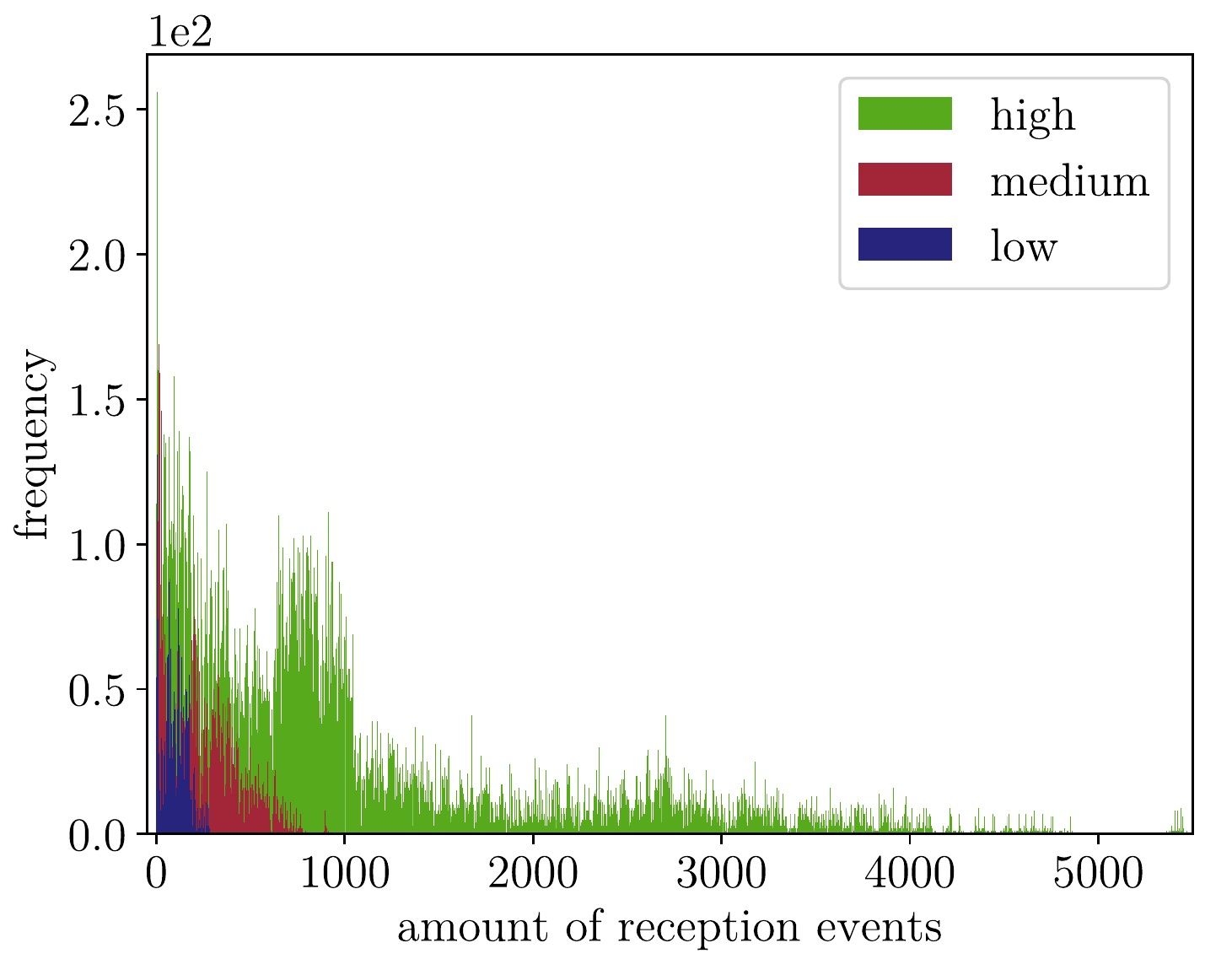}
  \caption{Histogram showing the raw amount of reception events in the simulations.}\label{fig:reception-event-density}
  }
  \hfill
  \parbox{0.495\textwidth}{
    \includegraphics[width=0.5\textwidth]{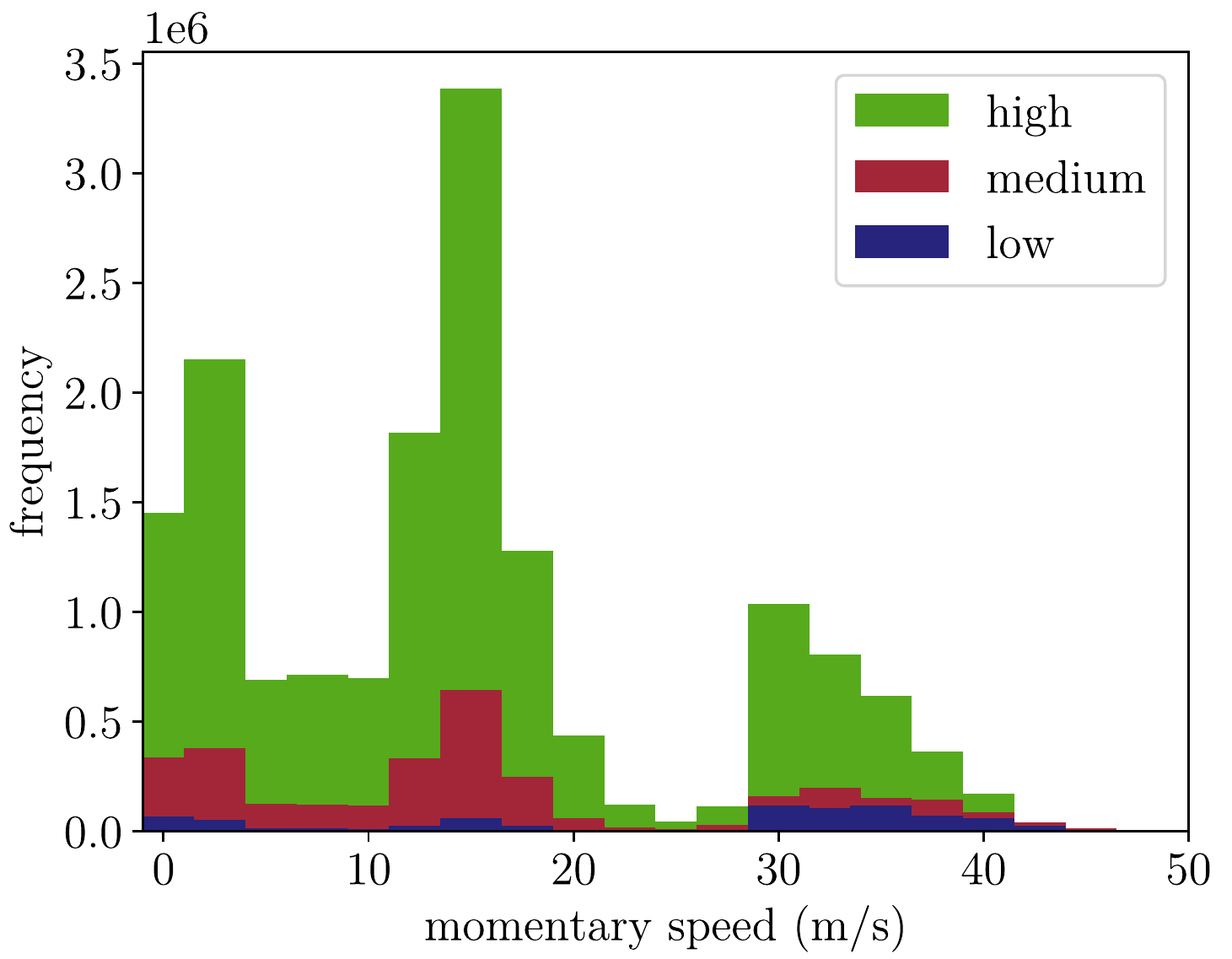}
  \caption{Histogram showing distribution of speed in the simulations.}\label{fig:speed-density}
  }

\end{figure}

The amount of messages transmitted in the simulations varies between the simulations and densities; at low densities, 908 to 1144 messages are sent, at medium densities, there are between 3996 and 4489, and at high densities, there are 20482 to 21878 messages sent.
The corresponding reception events are much more scattered; each vehicle at different densities can receive 0 messages (e.g., if they are not close to any other vehicles).
For low density, a vehicle receives up to 278 reception events (total over all low density simulations is 277916 events spread over 2820 receivers), while at medium density this number goes up to 911 reception events (total over all 1815215 spread over 7695 receivers).
Finally, for a high density, a single vehicle processes up to 5468 reception events in the 100 simulation seconds (total over all simulations over all 37500 vehicles is 37043160), or about 1000 messages per vehicle (10 per second, i.e., roughly 100 nearby vehicles at a beaconing rate of $1Hz$ if we ignore lost messages).
A graphical view of reception event frequency is given in the histogram in Figure~\ref{fig:reception-event-density}.

The scenario also includes a wide variety of traffic behavior, as illustrated in Figure \ref{fig:speed-density}, which shows aggregate speed statistics over all runs in a specific density.
The statistics were computed by taking the current local speed vector for every vehicle for every position update (which happens at $10Hz$) and aggregating all these samples.
This results in a mean speed of $24.36$ m/s for the low density scenario, with a very large standard deviation of $13.73$ m/s; since the median speed is $30$ m/s, this suggests that most of the deviation is due to traffic lights.
In the medium density configuration, the median ($13.33$ m/s) and mean ($15.06$ m/s) drop significantly, although the amount of vehicles in the simulation is fairly low (only about 2.5 times the vehicles compared to a low density); the standard deviation is still very high ($12.34$ m/s), indicative of the wide variety of driving behavior.
Finally, our high density scenario drops down further to a mean speed of $12.81$ m/s, with a standard deviation of $10.94$ m/s, while the median is $12.81$ m/s.

\subsection{Limitations}

Our dataset cannot be fully representative of all possible attacks in VANETs, especially because the implemented attacks are representative of a specific type of attack.
Investigating the effect of multiple attack types across a single simulation is not possible with this dataset.
We argue that our dataset should be used as a starting point for a more rigorous approach to the evaluation of such systems -- other researchers can use this process to find weaknesses in our detection approaches and implement new attacks.
We believe this process is essential to achieving scientifically meaningful results: existing work nearly always relies on non-published code in some way, and thus it is very difficult to verify others' results.
This leads to difficulty in replication of results, especially for complex detection systems that have many moving parts.
The purpose of this dataset publication is to alleviate this: authors can make verifiable and reliable comparisons between their schemes and ours.

Another important limitation of our dataset is that the evaluation workflow in Figure \ref{fig:workflow} is fundamentally non-interactive: it is designed for \emph{detection}, not for \emph{response}.
This means that some specific misbehavior detection schemes that rely on interactivity or application decisions based on the detection of an attack (e.g., increasing safety distance in autonomous driving) cannot be evaluated with our dataset directly.
However, for systems that protect specific applications, a comparison with other schemes always requires custom implementation.
The core weakness of our approach is that we cannot directly evaluate trust over time without major modifications to our workflow, since trust schemes often do not output decisions for every message.

\section{Metrics}
\label{sec:metrics}

Detection performance is a complicated and multi-faceted issue, whose definition also varies across publications, depending mostly on the purpose of the detector.
Even in intrusion detection in general, determining how to evaluate detection mechanisms and how to choose the appropriate mechanism for deployment, is considered a challenging issue, and the trade-off is non-trivial~\cite{Cardenas2006}.
In misbehavior detection specifically, many authors use false positive/negative rates or equivalent metrics to determine how well attacks are detected, and this is combined with other performance metrics (such as latency, or application-specific metrics).
Although these metrics are useful to compare performance of mechanisms, we find that there is a lack of metrics that are useful for the development of new detectors.
In this section we propose an additional metric that fills this gap.

Another issue that should be addressed is specific to detection in \emph{distributed} peer-to-peer systems: how should detection metrics be aggregated across participants?
For example, given a simulation with two honest vehicles and one attacker, how do we characterize the detection performance of the same detection system (running within two vehicles independently)?
We previously touched on this issue in a discussion with the vehicular communication community~\cite{vanderHeijden-fg}.
In this paper we will quantify the detection quality by classifying every detection event as true/false positive or as true/false negative; a detection event occurs whenever a message is received (i.e., we assume the detection decision is made as soon as possible after reception).
We aggregate these results by counting the errors generated by \emph{detection events}, not in terms of \emph{sent messages} or \emph{participants}.
Which aggregation method is chosen is highly relevant for the interpretation of the results: in this work we focus on detection events to obtain a picture of the overall quality of the results.
Aggregating by \emph{message} provides information about how well a specific message sent by the attacker is detected, but presenting the results in terms of detected messages would mean that the amount of receivers is completely disregarded.
Similarly, aggregation by \emph{participant} ignores how much contact this vehicle has with the other vehicles.
Since the amount of messages between vehicles is also indicative of a potential impact of an attack, aggregation by detection events is the best approach for an overall evaluation of detector performance.
However, we point out that these metrics can also be implemented with our dataset, since we provide message and sender labels for every message.

\subsection{Evaluating Detection Quality}

The first metric we use to decide the quality of the detector is based on the well-established confusion matrix (which basically corresponds to an overview of true/false positives ($TP/FP$) and true/false negatives ($TN/FN$)).
There are many options to choose from here; for example, \emph{accuracy}, defined as the number of correct classifications ($TP+TN$) over all classifications ($TP+FP+TN+FN$), appears intuitive but suffers from the accuracy paradox for imbalanced sets.
It is thus considered good practice~\cite{Davis2006-PR-ROC,Saito2015-PLoSONE-PR} to always provide a quantification with two values, showing the trade-off between increased false positives to reduce false negatives and increased false negatives to reduce false positives.
One such formulation is the use of \emph{precision} and \emph{recall}: precision quantifies the relevance of detection events ($TP/(TP+FP)$), while recall quantifies what rate of positives is actually detected ($TP/(TP+FN)$).
An optimal detector thus has a precision and a recall of $1$; how significant a deviation from this value is acceptable depends strongly on the application.

The state of the art~\cite{vanderHeijden-survey} typically reports false positive ($FP/(FP+TN)$) and true positive ($TP/(FN+TP)$) rates, which provides a different and significantly skewed picture in certain situations, as discussed in machine learning literature~\cite{Davis2006-PR-ROC,Saito2015-PLoSONE-PR}.
Specifically, precision and recall are more informative in situations where a binary classification task (e.g., packet maliciousness decisions) is performed on an imbalanced dataset.
As our dataset contains attackers in different degrees, and the amount of decisions made for attacker-transmitted messages ($TP+FN$) compared to the amount of decisions made for benign messages ($FP+TN$) is significantly different, we should thus prefer precision recall curves.
As pointed out by other authors~\cite{Davis2006-PR-ROC}, a detector that is better in the PR graph is guaranteed to be better in the ROC graph; the interpretation process is generally similar (i.e., which curve is closer to the optimal point).

PR graphs provide us with an overall estimation of detector performance, but they have an important disadvantage: they are generally not as easy to interpret as a graph with FPR/TPR (referred to as an ROC curve).
This greatly impacts the use of PR graphs in the literature: not only are they somewhat harder to understand fully, PR graphs often ``look much worse'', as demonstrated by Davis~\&~Goadrich~\cite[Fig. 1]{Davis2006-PR-ROC}.
This figure shows that the ROC curve can look close to optimal (the \emph{area under curve} (AUC) is large), while the PR curve for the same data looks much worse (the AUC is small).
This is partially related to the fact that interpolation between points on a PR curve is non-trivial; for details, refer to~\cite{Davis2006-PR-ROC}.
In addition to this issue, PR-graphs do not provide information about where potential flaws of individual mechanisms are, or whether a combination of multiple detectors can out-perform the individual mechanisms (as we argue in previous work~\cite{Dietzel2014-smartvehicles}).

\subsection{Evaluating Detector Limitations}

To study the limitations of detectors without arbitrarily guessing which factors may influence such detectors, we design a new metric to find indications of such influences.
The idea of our metric is quite simple: examine whether the distribution of erroneous classification rates (i.e., false positives and false negatives) is uniform over the receiving vehicles.
If this metric says the distribution is uniform, the detector performance is not dependent on factors that are varied in the simulation, such as which vehicle executes it, or the relative position between the receiving and sending vehicle.
On the other hand, if this distribution is extremely skewed, the conclusion is that the detector performance depends a lot on the context of the vehicle.
We expect this is the case for many misbehavior detection schemes (and indeed most of the literature just assumes this is true), but it is also valuable information to know where the discrepancies occur.
This enables further investigation into the detector's strengths and weaknesses, and finding a skewed distribution would suggest that combining the results of different mechanisms is the way forward.
Note that this is \emph{not a qualitative metric}: uniformly good or poor error dispersion does not imply that a metric is significantly better or worse, it only suggests whether there is room for improvement.

Given this intuition, we investigated and found a metric for statistical dispersion that is commonly used in sociology and economics to measure income inequality: the Gini coefficient or Gini index~\cite{GiniCoefficient}, originally defined in 1987 by Dixon~et~al.
The idea can be visualized by sorting people by income in ascending order and then plotting the cumulative fraction of this list against the cumulative income of that group.
More formally, the Gini index $G$ of a population with mean size $\mu$ and value $x_i$ assigned to individual $i$ is defined as:
\begin{equation}
  G = \frac{\sum^n_{i=1} \sum^n_{j=1} \left| x_i - x_j \right|}{2 n^2 \mu}
\end{equation}

The Gini index itself is not novel, nor is the application to quantify errors (see e.g.~\cite{Eberz2017}), but our application of it is slightly different: we propose that it can be used effectively to determine the statistical dispersion in the error rates across vehicles.
The reasoning is that computing the overall performance as discussed in the previous section hides localized effects associated with individual vehicles.
Thus, if a mechanism has some regions where it performs really well (e.g., a highway), while it performs very poorly in other regions (e.g., urban settings with lots of traffic lights), these effects will be averaged out in the overall performance.
If the overall performance is reasonable, one can use the dispersion in the error rates to determine whether this happens both for false positives and for false negatives (the latter being dependent on the attack): the higher the Gini index of these rates, the more differences exist between vehicles.
However, if performance is poor overall, the Gini index can still be close to zero (or conversely, be close to 1); the arrays $(0.1, 0.1, 0.1, 0.1)$ and $(0.9,0.9,0.9,0.9)$ have the same Gini index of zero.
There are at least two main ways to use the result of this metric: 1) investigate the vehicles on either side of the skew and see whether the detector can be improved by changing its' functionality or 2) investigate whether fusion can be used to exploit low amounts of errors produced by different detectors in different scenarios.

\section{Evaluation of Plausibility Detectors}
\label{sec:detection}

This section shows an application of our dataset and metrics to analyze several data-centric plausibility detectors, which are detectors that verify a received message against data from local sources only.
The decisions made by these detectors are practically instant (i.e., they do not depend on other data sources), and it does not generate additional attack vectors that can be used for \emph{bad mouthing} and similar attacks, as is a risk in trust schemes and consistency mechanisms~\cite{vanderHeijden-survey}.
As plausibility mechanisms are often used as a basis for trust establishment~\cite{LeinmuellerART,Raya-INFOCOM-2008,SchmidtVEBAS,Stuebing-Kalman}, we focus on these.
We implement detectors in our detection framework Maat\footnote{\url{https://github.com/vs-uulm/Maat}}, which is a detection and fusion framework based on subjective logic that we are currently developing.
In this work, we compare four: the \emph{acceptance range threshold} (ART), the \emph{sudden appearance warning} (SAW), the \emph{simple speed check} (SSC), and the \emph{distance moved verifier} (DMV).
Of these, the acceptance range threshold is the most well-studied, originally introduced by Leinm{\"u}ller~et~al.~\cite{LeinmuellerART} and later used by others, including St{\"u}bing~et~al.~\cite{Stuebing-Kalman} and in our earlier work~\cite{vanderHeijden2016-vtcfall}.
It basically uses the expected reception range as a measure for the plausibility of the position included in incoming single-hop beacon messages, which are the most important source of information for VANET applications.
The sudden appearance warning was also introduced by Schmidt~et~al.~\cite{SchmidtVEBAS}, and is based on the assumption that vehicles will not suddenly appear, but rather always approach from a distance; if a message originates close by with an unknown sender, it is considered malicious.
The simple speed check and distance moved verifier were implemented as part of our work on a detection framework, and both examine whether a new beacon confirms information claimed in an older beacon.
The simple speed check decides maliciousness based on how the claimed speed relates to the speed implied by the position and time differences between the current and the previous beacon, and the claimed speed in the current beacon.
If the deviation exceeds a threshold, this detector classifies the message as malicious (similar to, but much simpler than, a Kalman filter~\cite{Stuebing-Kalman}).
Finally, the distance moved verifier checks whether the vehicle moved a minimum distance (similar to the way the MDM proposed by Schmidt~et~al.~\cite{SchmidtVEBAS}), and if this distance is too small, the message is considered malicious.

This selection of mechanisms is made for several reasons: 1) all of these mechanisms are exceedingly simple, 2) these mechanisms are designed to detect false positions in some sense, but as our analysis will show, different mechanisms detect different attacks, 3) the mechanisms rely on different data elements in the packet.
Especially the simplicity is important for this discussion, since this allows us to not only compare the mechanism performance dependent on their respective thresholds, but also showcase how our metrics and dataset enable a useful and detailed analysis of mechanism behavior.
We also focus on position verification as a specific application, in order to focus on a specific set of attacks, as discussed previously.
Finally, these are the mechanisms for which the source code is available, unlike other mechanisms we have found in the literature -- re-implementing mechanisms can be challenging, as often the implementation details are missing due to space limitations, and the code is not publicly available.

\begin{table}
  \centering
  \caption{Detector parameters, chosen based on earlier work.}\label{tab:detector-params}
  \begin{tabular}{l |l| l}
    Detector & Parameter & Values \\ \hline
    ART & est. reception range (m) & $100, 200, 300, 400, 450, 500, 550, 600, 700, 800$\\
    SAW & max. appearance distance (m) & $25, 100, 200 $ \\
    SSC & max. speed deviation (m/s) & $2.5, 5, 7.5, 10, 15, 20, 25$ \\
    DMV & min. distance (m) & $1, 5, 10, 15, 20, 25$ \\
\end{tabular}
\end{table}

\subsection{Results: Detection Performance}

\begin{figure}
\centering
  \parbox{0.495\textwidth}{
    \includegraphics[width=0.5\textwidth]{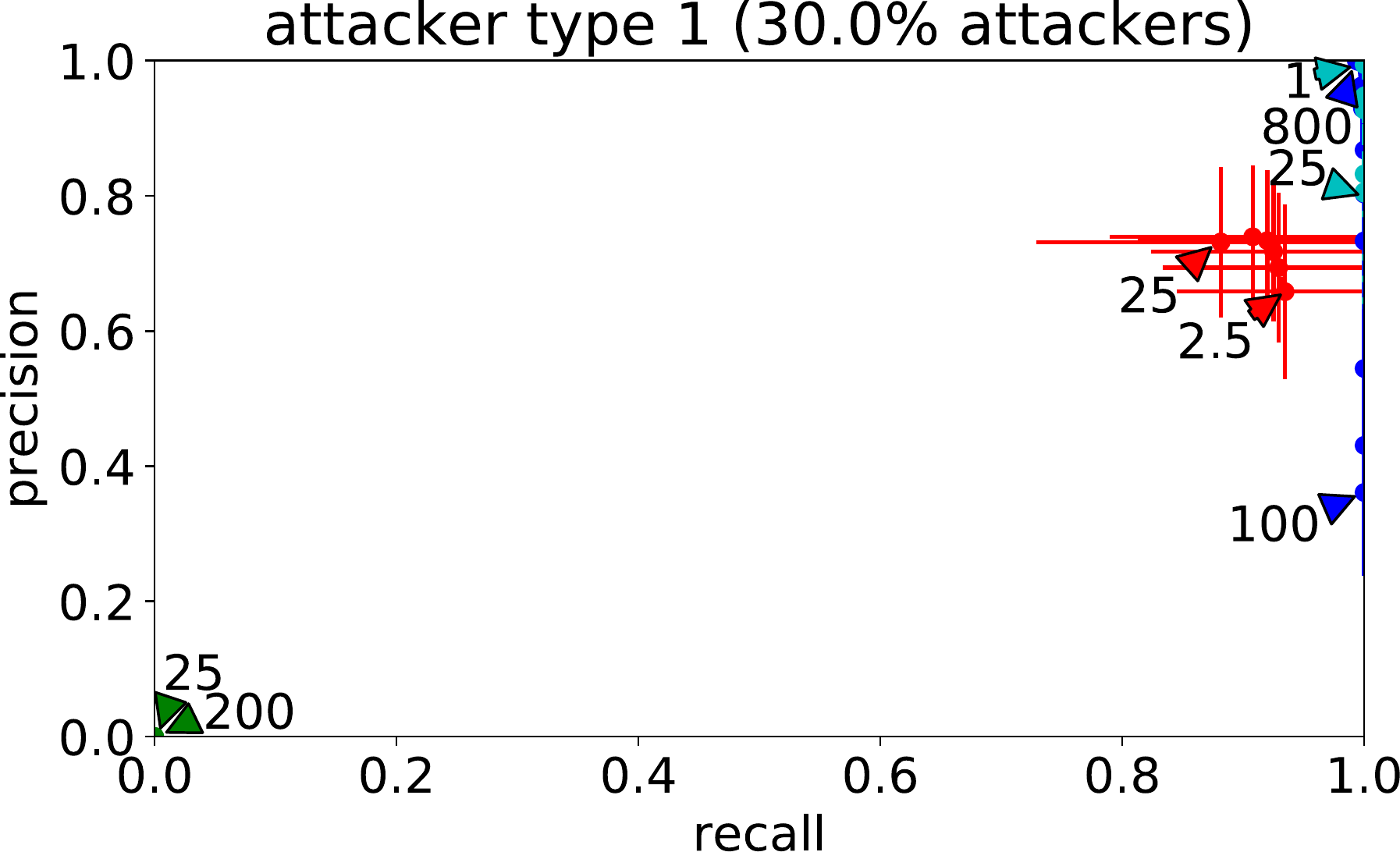}
  }
  \hfill
  \parbox{0.495\textwidth}{
    \includegraphics[width=0.5\textwidth]{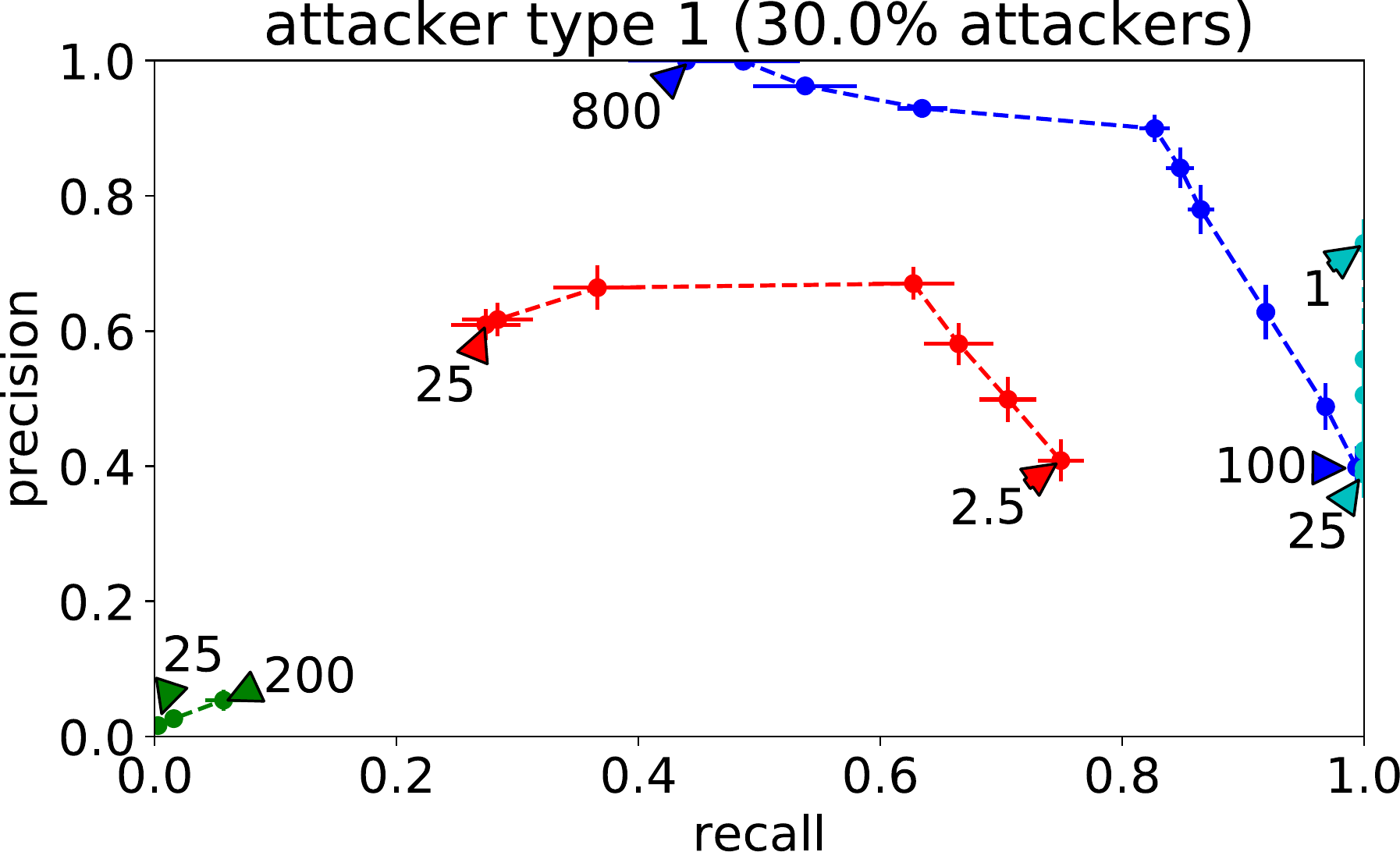}
  }

\centering
  \parbox{0.495\textwidth}{
    \includegraphics[width=0.5\textwidth]{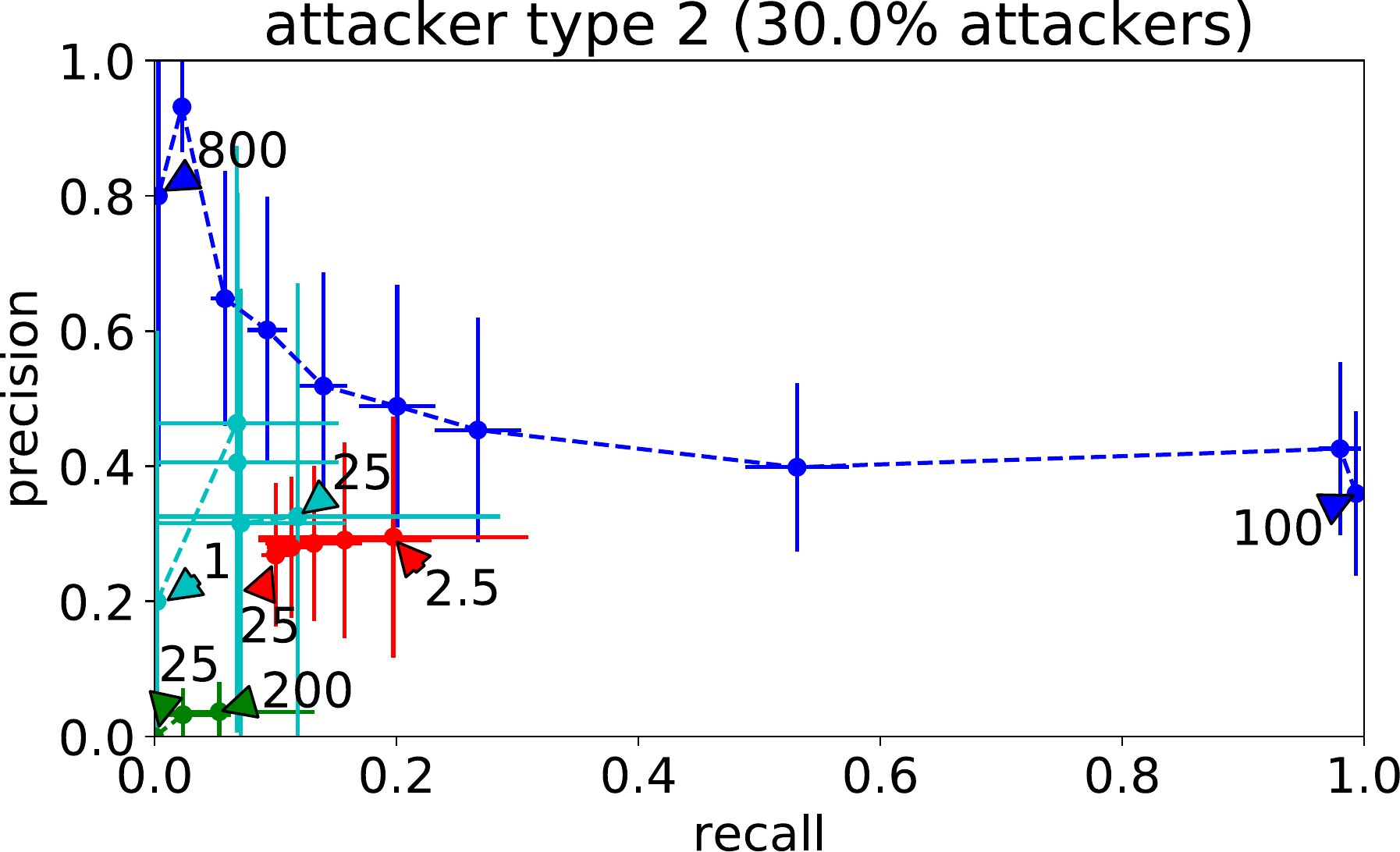}
  }
  \hfill
  \parbox{0.495\textwidth}{
    \includegraphics[width=0.5\textwidth]{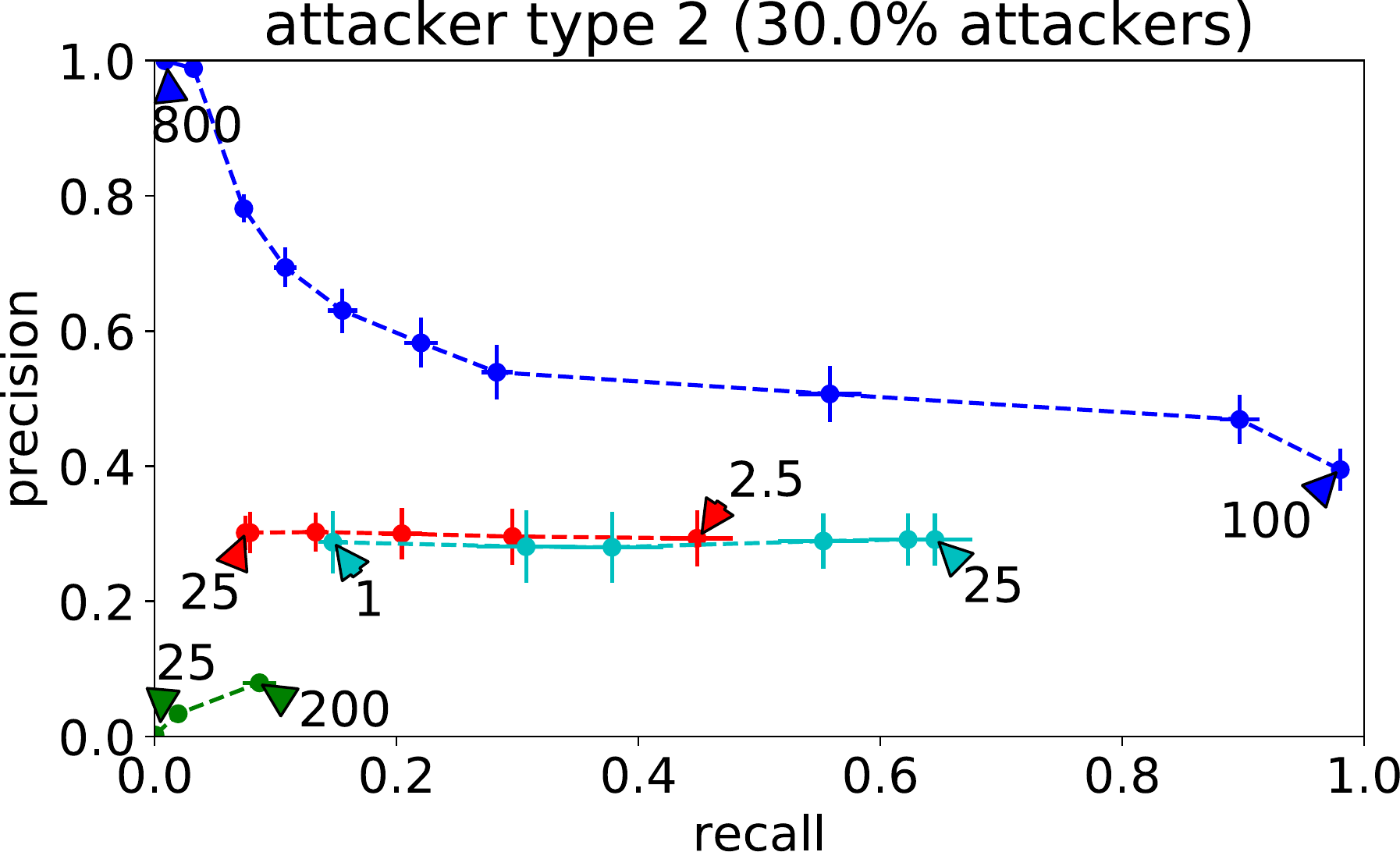}
  }

\centering
  \parbox{0.495\textwidth}{
    \includegraphics[width=0.5\textwidth]{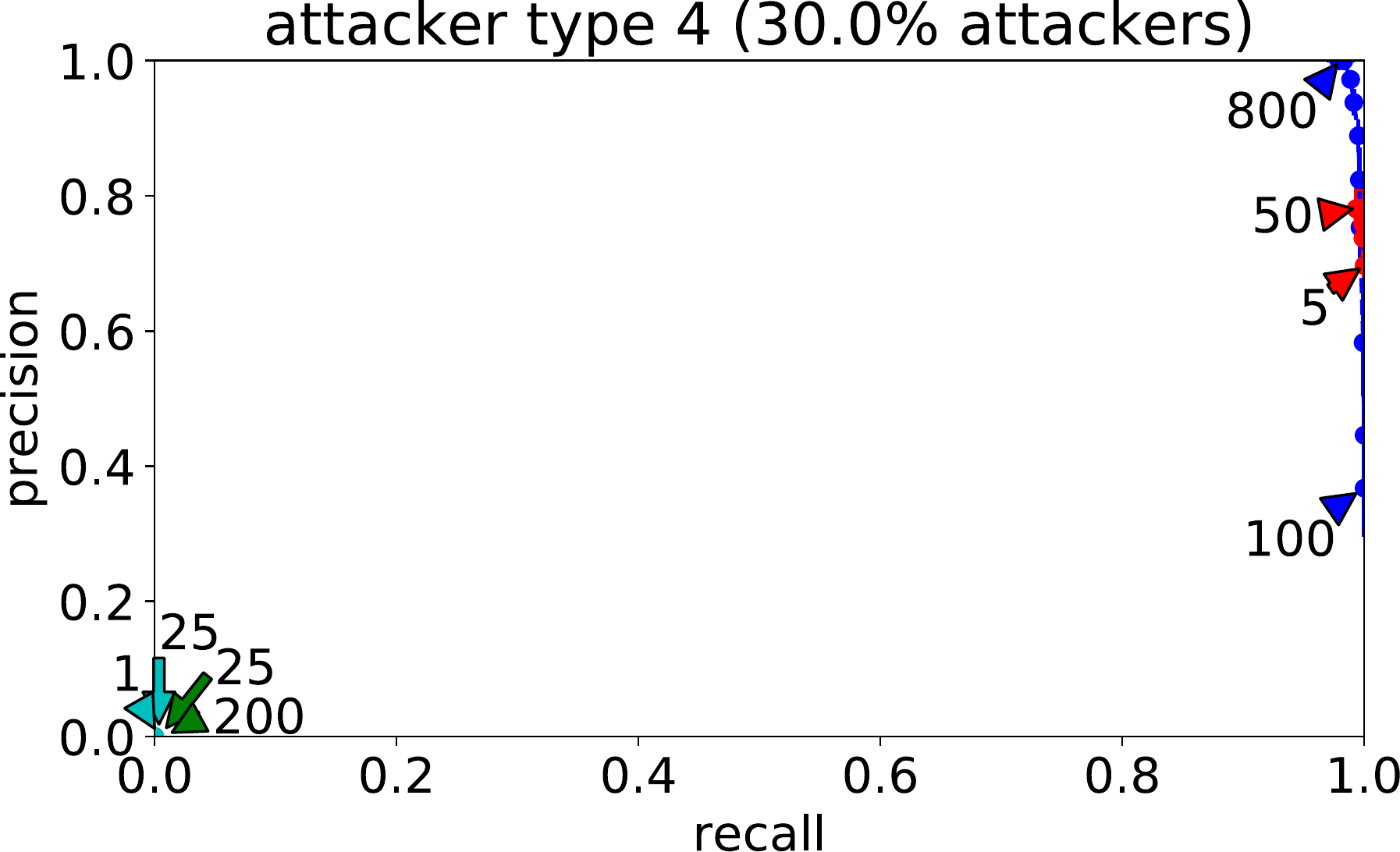}
  }
  \hfill
  \parbox{0.495\textwidth}{
    \includegraphics[width=0.5\textwidth]{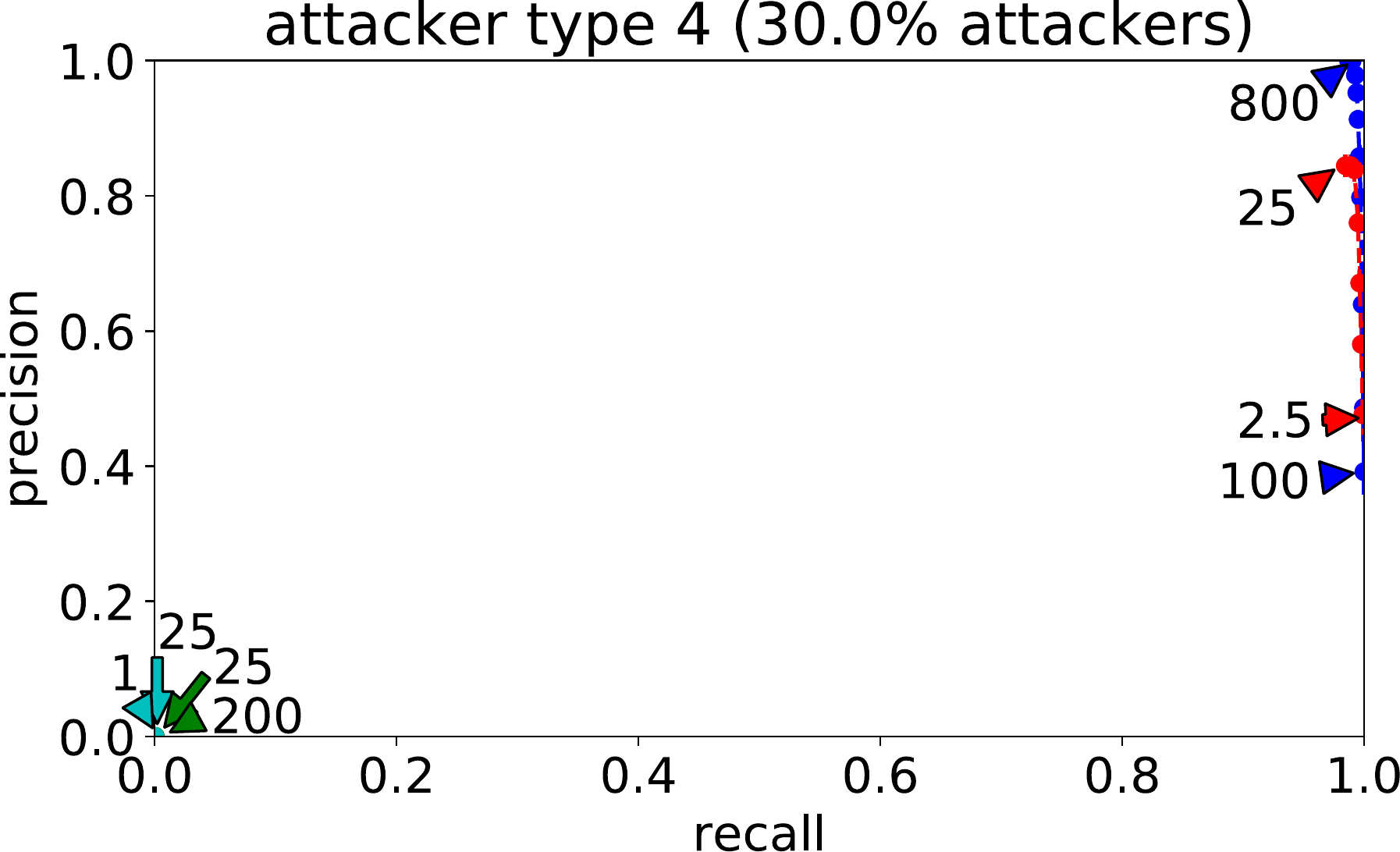}
  }

\centering
  \parbox{0.495\textwidth}{
    \includegraphics[width=0.5\textwidth]{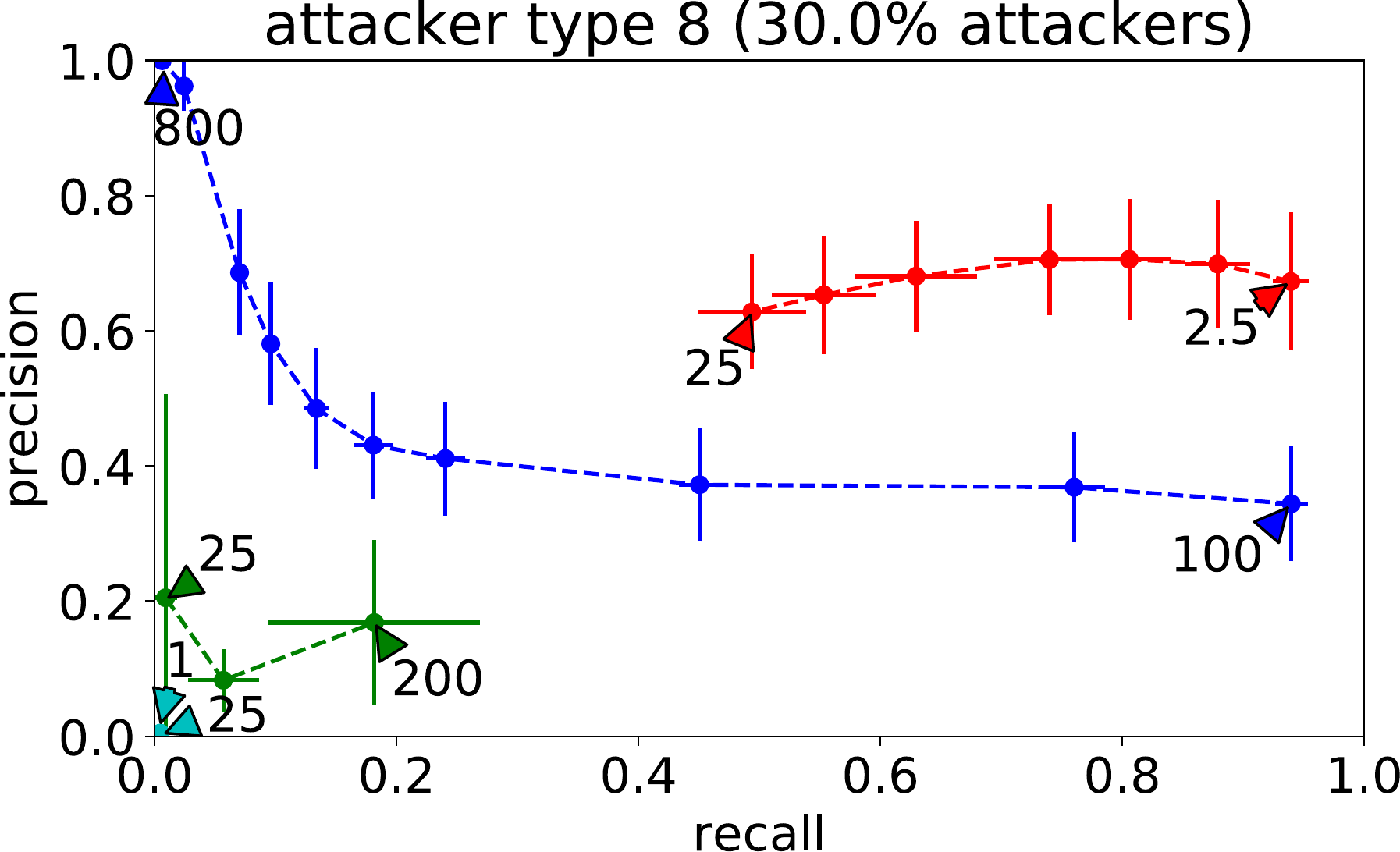}
  }
  \hfill
  \parbox{0.495\textwidth}{
    \includegraphics[width=0.5\textwidth]{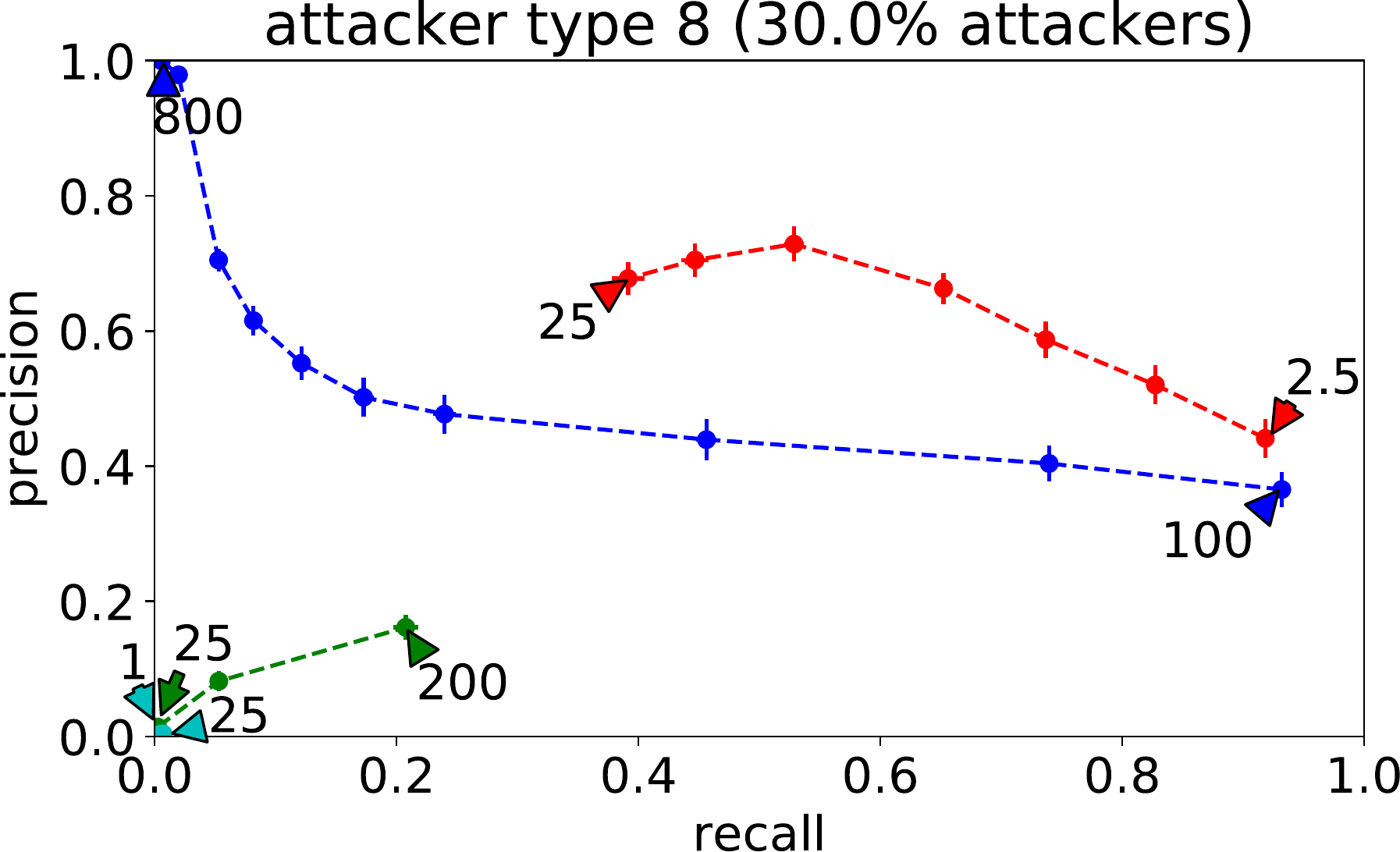}
  }

\centering
  \parbox{0.495\textwidth}{
    \includegraphics[width=0.5\textwidth]{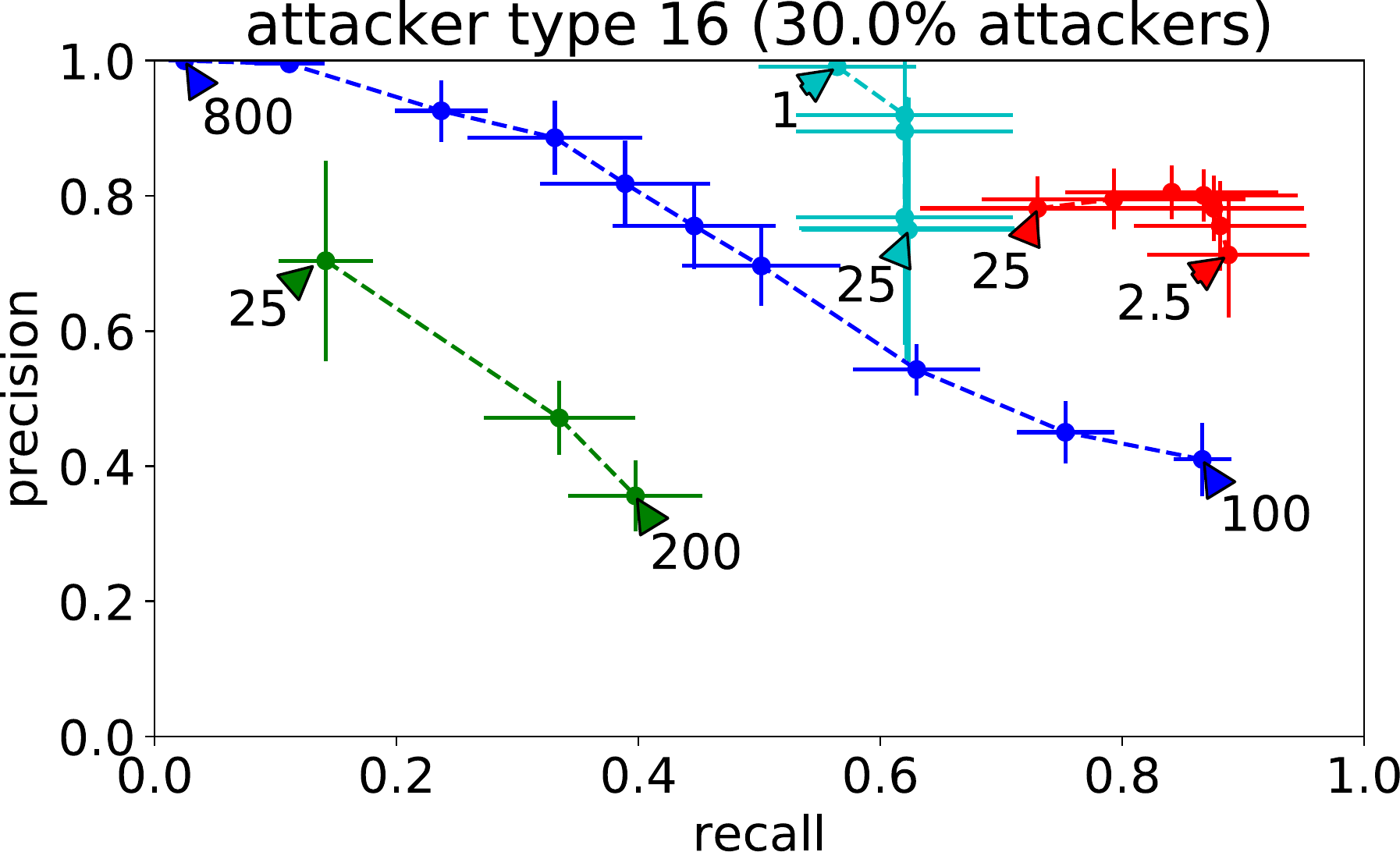}
  }
  \hfill
  \parbox{0.495\textwidth}{
    \includegraphics[width=0.5\textwidth]{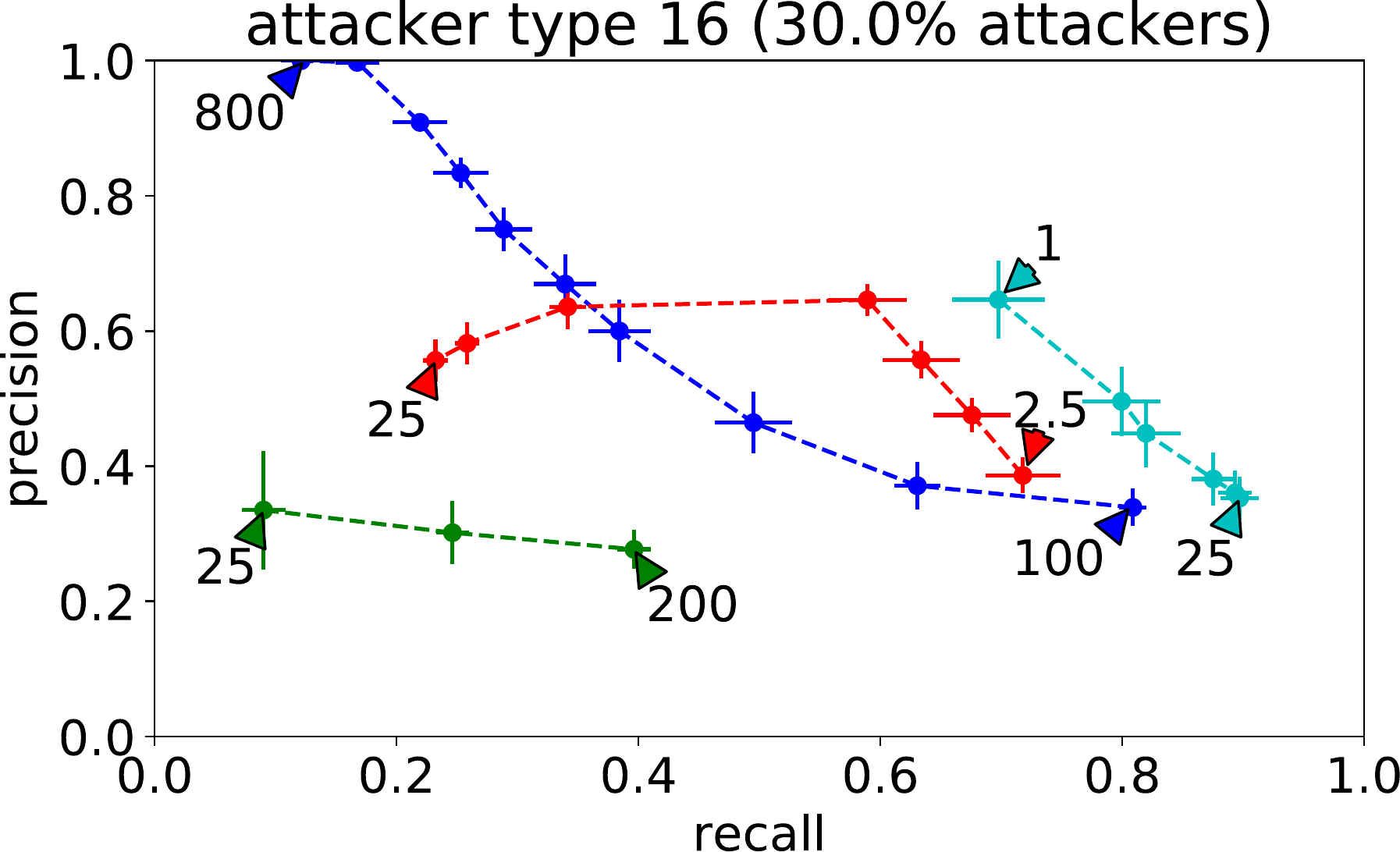}
  }

  \vspace{1em}

  \includegraphics[width=0.5\textwidth]{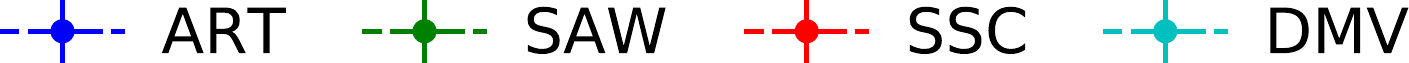}

  \caption{Precision-Recall graphs for low densities (left) and high densities (right).}\label{fig:pr-graphs}
\end{figure}

Here we show the analysis results of our misbehavior detection framework, Maat, executing the detectors described above with different parameters, as listed in Table \ref{tab:detector-params}.
Maat, which is currently in development within our institute, uses a graph representation to represent the data received by a vehicle, and is able to execute multiple detectors with multiple thresholds in parallel.
In this paper, we focus on the outcomes of individual detection mechanisms: for a real-world deployment of Maat, this evaluation process is the first step.
These results can be used to configure initial thresholds for each detector.
For brevity, we focus on high and low density scenarios with high amounts of attackers (30\%), as these provide the most notable output; we publish the entire set of figures and the underlying data as additional material\footnote{\url{https://github.com/vs-uulm/securecomm2018-misbehavior-evaluation}}.
The 10\% and 20\% attacker cases show comparable result for each set of graphs in Figure \ref{fig:pr-graphs}; similarly, the medium density is comparable to the high density (as the application behavior is quite similar, as illustrated by Figure \ref{fig:speed-density}).

As our dataset contains five simulation runs per behavior/attacker parameter set, each point in the graphs represents the mean of five runs, aggregated over vehicles as described previously.
The error bars in these graphs show the sample standard deviation associated with this mean.
The colors show the different detectors, also listed in the legend on the bottom; for black-and-white readers, we point out that the extremes of the threshold values (indicated with arrows at the extreme ends of each plot) are unique.
Finally, note that the lines in these graphs are for illustrative purposes only -- as previously discussed, interpolation between these points is a non-linear task~\cite{Davis2006-PR-ROC}.

\subsection{Discussion: Detection Performance}
In Figure \ref{fig:pr-graphs}, the different attackers are listed from top to bottom as specified in Table \ref{tab:attacks}.
Overall, as one might expect, the type of attack is an important distinguishing factor in the effectiveness of the detection process (i.e., easily detected attacks generally have higher recall).
One can observe immediately that the results for the different detectors vary greatly per attack, regardless of density.
However, some detectors' performance is dependent on the density of the traffic (DMV is notable here).
It can also be seen that the SAW has very poor performance for all attacks except for 16 -- this corresponds to expectations from the detector design.
We now focus on a brief discussion of each individual attacker.

For the first attacker, which falsifies position, the results show that in low density settings, all detectors except SSC very accurately detect the attack.
A similar trend can be observed in the high density scenario; however, note that the ART performs slightly worse at very high thresholds (greater than 500); this conforms with results from an earlier study~\cite{vanderHeijden2016-vtcfall} (with different data and a different implementation).
With regards to the SSC, which verifies whether the claimed speed in the current beacon corresponds to the distance moved between two beacons.
However, note that this is not necessarily correlated with an attack: it occurs naturally in the application behavior that vehicles' speed deviates significantly from the movement, for example when breaking for traffic lights.
Since no interpolation is performed by the SSC based on other information (such as sensor measurements) and the beacon frequency is relatively low, this mechanisms' performance is overall quite poor.
Note that vehicles do drive by the position claimed by the attacker (i.e., the attack position is within the scope of the simulation).

The second attacker, type 2, adds a fixed vector to its' position; this attack is harder to detect for most mechanisms, and this can be observed by the poor performance in all cases.
The very large standard deviation in the low density case (left) suggests that the success is very dependent on the relative position of the vehicles; especially for ART, this is exactly what one would expect.
This is confirmed by the greatly reduced deviation observed in increased densities.
Since the attacker adds exactly the same value to each beacon, it is expected that the DMV does not perform at all: indeed, this effect can be observed very well in the high density graph (precision remains constant at 0.3, the attacker fraction, for all thresholds).
A very similar behavior is shown by the SSC; again, this is expected, since the relative position claimed by the attacker is the same as the ground truth.

Attacker type 4, which transmits a random position from the simulation area (essentially corresponding to a broken GPS), is never detected by the DMV (since the probability that two positions near the same area are chosen is very close to zero).
The ART and the SSC have no problems with this attacker, which is quite easy to detect.
It can, however, be observed that low ART thresholds result in a low precision.
Note that in this case, the randomness introduced into the attack results in a poorer performance; this attacker fits more to faulty behavior than to an attack (which is commonly also classified as misbehavior~\cite{vanderHeijden-survey}).

The next attacker, attacker type 8, shows remarkably similar behavior to attacker 2 for the mechanism (again, confirming previous results~\cite{vanderHeijden2016-vtcfall}).
However, due to the randomness in this attacker, the attacker is somewhat harder to detect for the ART than before, and cannot be detected at all by the DMV.
The SSC, on the other hand, appears to be surprisingly suitable for this attack.
This information suggests fusion may be a suitable option to investigate in future work.

Finally, our last attacker (attacker type 16) is different from the previously discussed attacks, in that it changes the vehicles' messages in a pattern over time (as opposed to manipulation of individual messages independently, as done previously).
This is noticeable in the very different detection behavior, in particular of the DMV, since the attacker is essentially converging to a situation where they do not move at all (which the DMV easily detects).
However, ART and SSC behavior is comparable to attackers 8 and 1 -- as expected: the attack could be seen as a transition from attacker 8 to attacker 1 over some time.

In summary, we can conclude that the ART with a high threshold works well against attackers that transmit erroneous positions (attackers 1 and 4), but has significant difficulties with those that are designed to confuse applications (attackers 2, 8 and 16).
Against these malicious cases, the SSC works surprisingly well with lower thresholds, but it is subject to very poor performance against attacker 2.
The DMV mechanism works best in dense traffic against attacker 16, and it also does well against attacker 1, but overall its' performance is very poor: this mechanism is clearly only suitable to identify very specific attacks.
We also note that the SAW does not outperform any mechanism in any scenario; a future study that includes ghost vehicles (similar to for example, \cite{Lo2007}) could show some benefit, but the extremely low precision will require some effort to make this scheme deployable.
Finally, note that ART and SSC appear to out-perform each other depending on the scenario (and the configured threshold): these are mechanisms we will focus on for our examination of the dispersion.

\subsection{Results: Dispersion of Errors}

\begin{figure}
\centering
  \parbox{0.495\textwidth}{
    \includegraphics[width=0.5\textwidth]{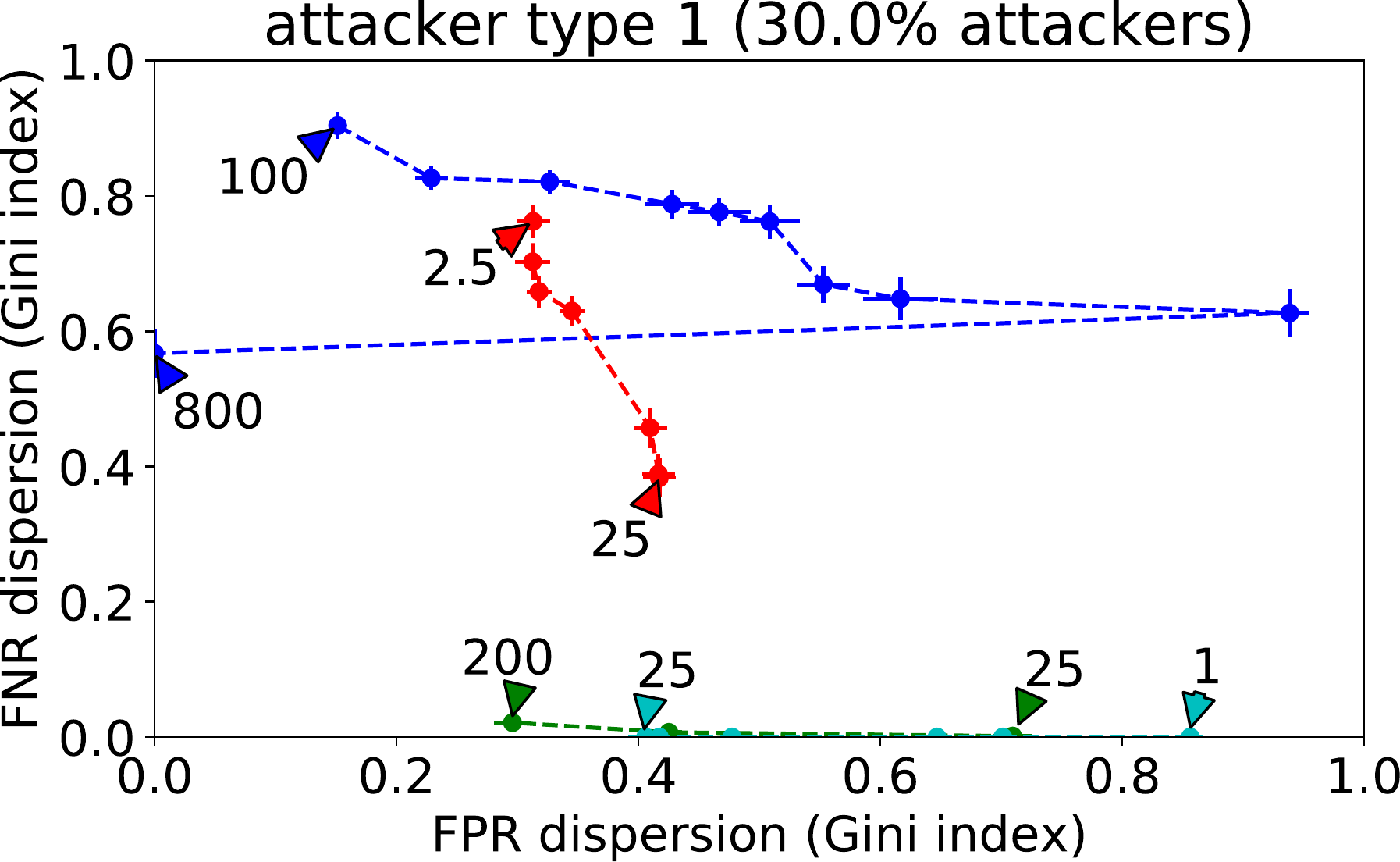}
  }
  \hfill
  \parbox{0.495\textwidth}{
    \includegraphics[width=0.5\textwidth]{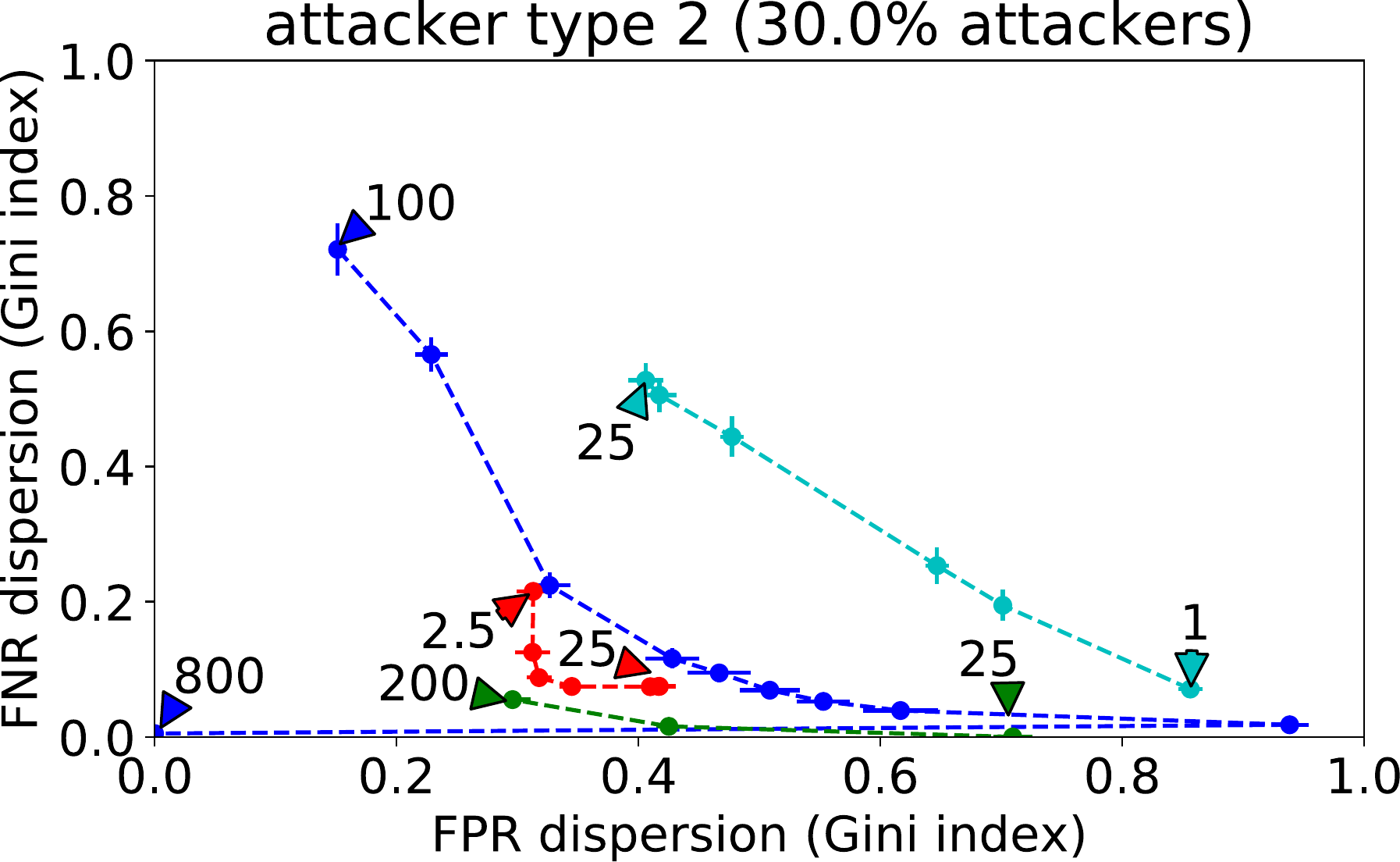}
  }

  \parbox{0.495\textwidth}{
    \includegraphics[width=0.5\textwidth]{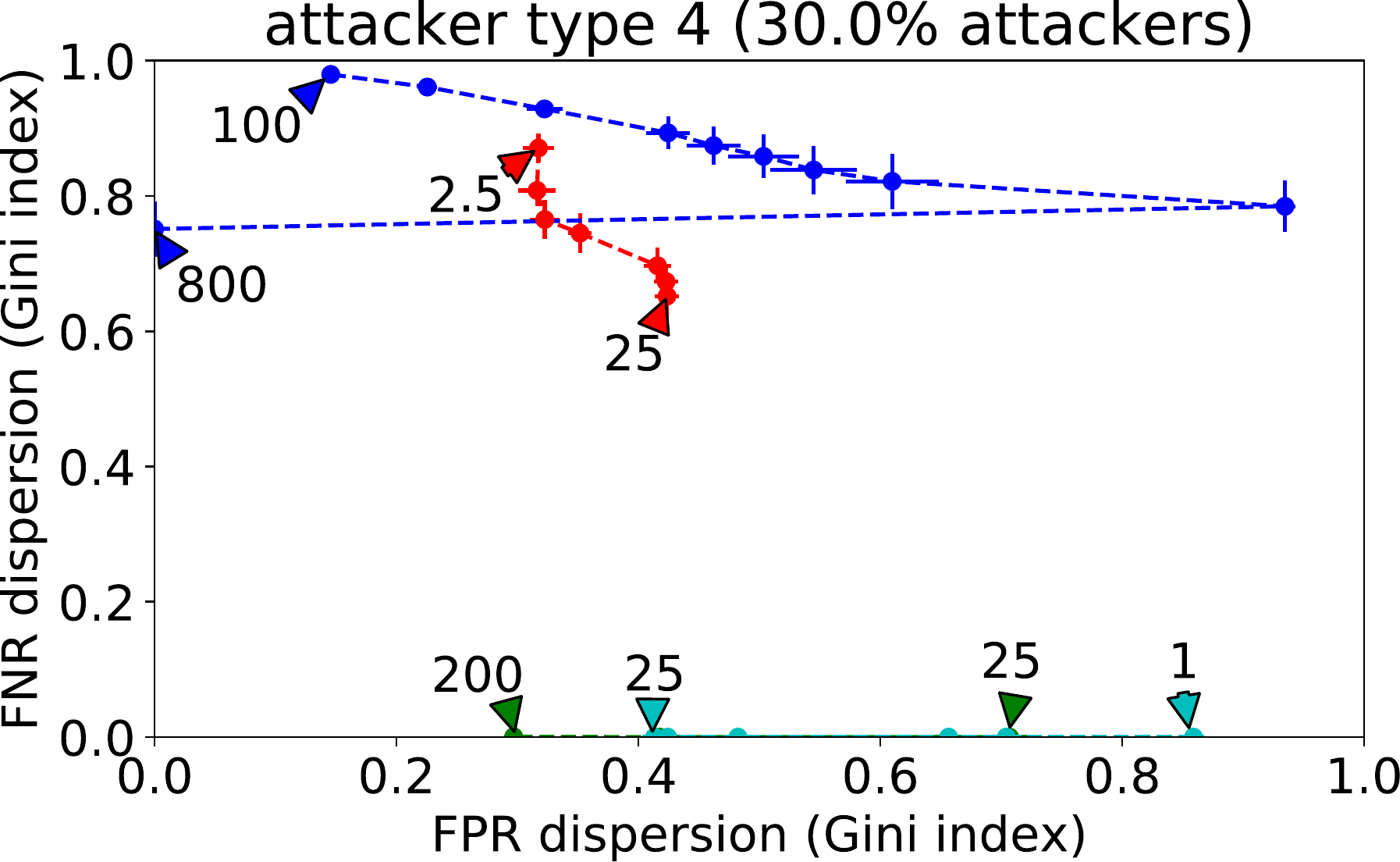}
  }
  \hfill
  \parbox{0.495\textwidth}{
    \includegraphics[width=0.5\textwidth]{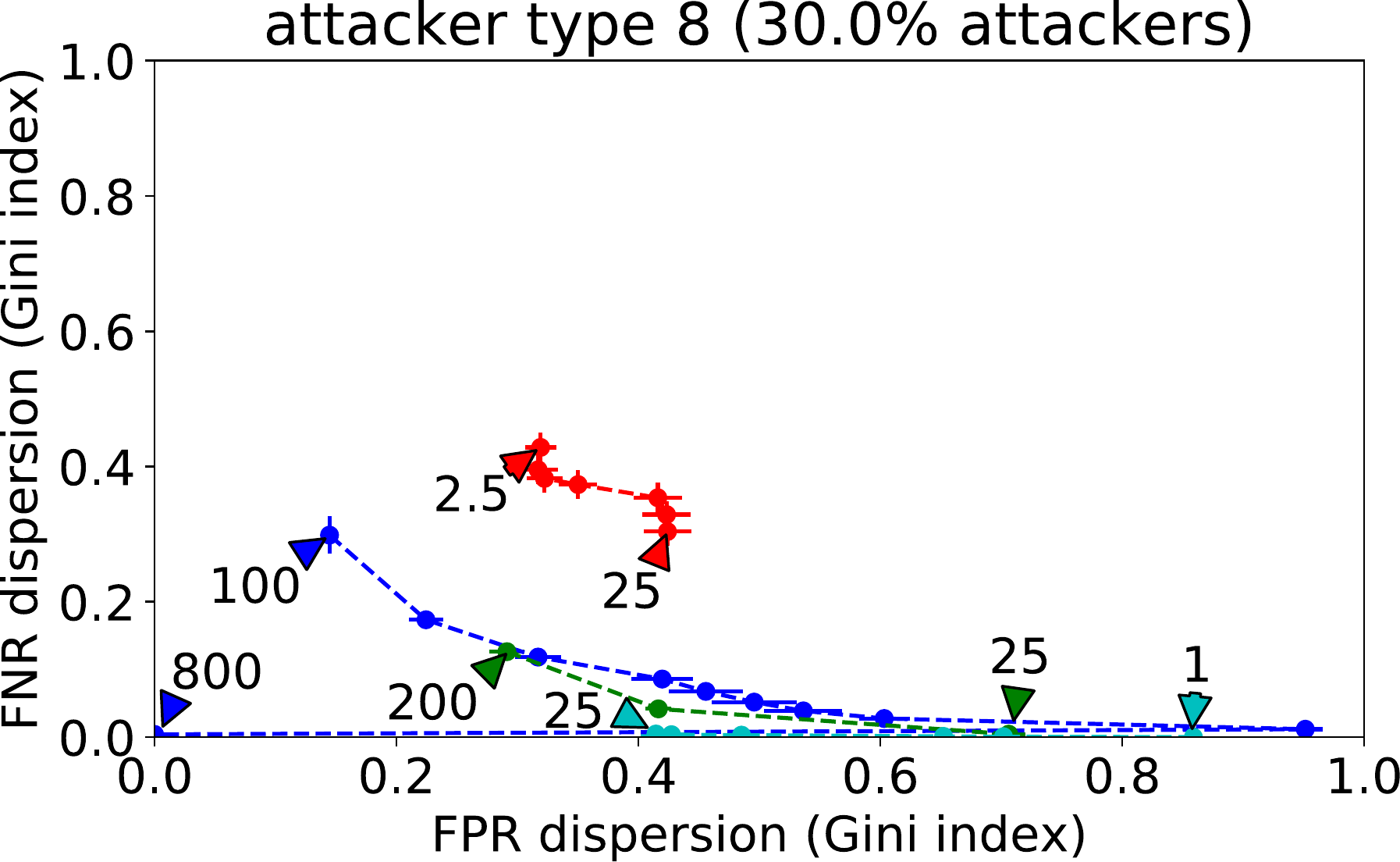}
  }

  \parbox{0.495\textwidth}{
    \includegraphics[width=0.5\textwidth]{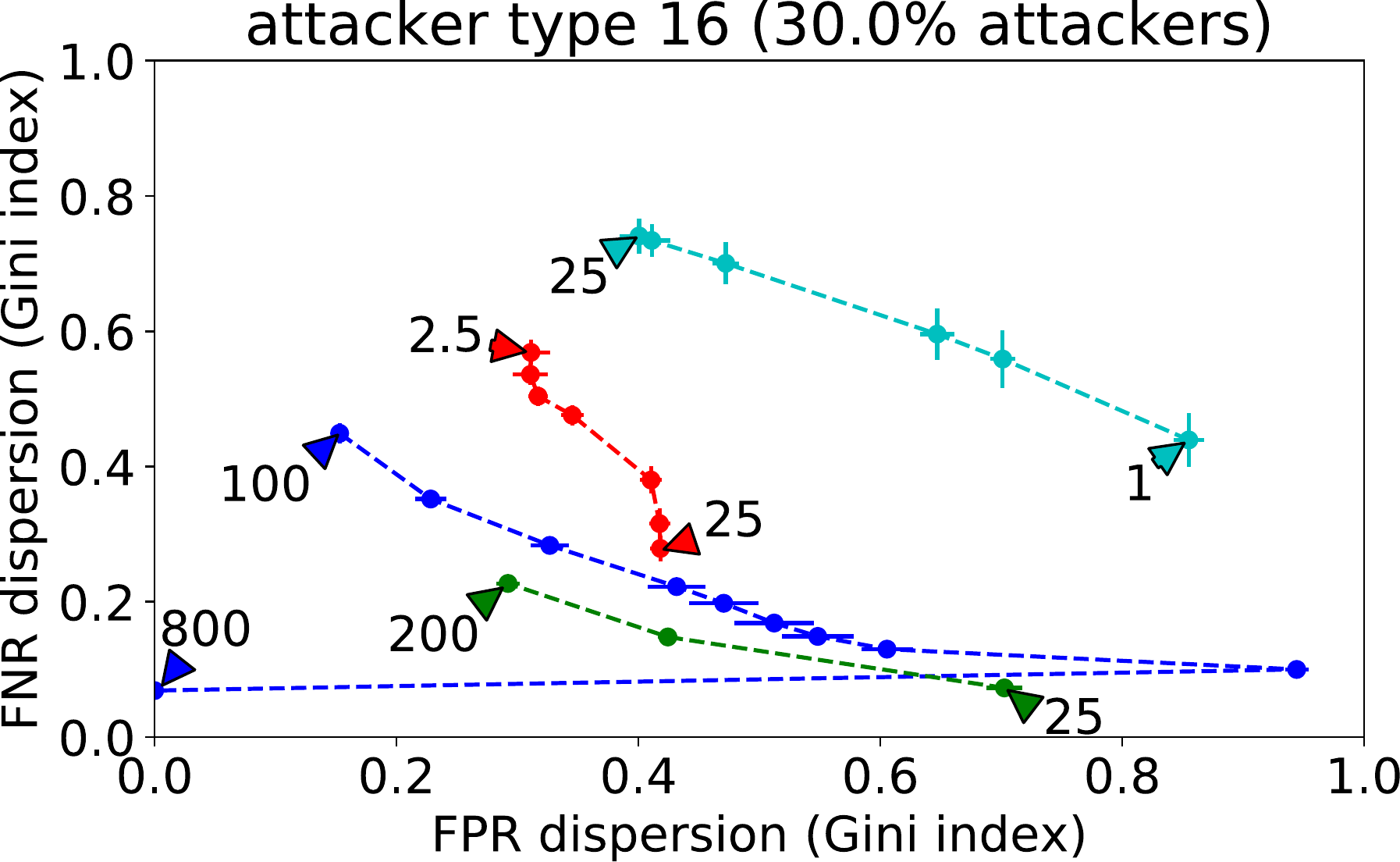}
  }

  \vspace{1em}

  \includegraphics[width=0.5\textwidth]{img/legend.pdf}

  \caption{Gini indices of FPR and FNR for different attackers}\label{fig:attacks-gini}
\end{figure}

Now that we have reviewed the detection performance in terms of precision-recall (PR), we examine our new metric based on the Gini index to study how to improve detection performance.
Preliminary analysis has shown that the Gini index is not meaningful for small sample sizes as in our low density results, since the population is too small to make meaningful statements (because the sample standard deviation is very large for these results).
The data and graphs are available for future analysis, but we caution against drawing conclusions from these for this reason.
We therefore focus on a discussion of the high density scenario, which contains enough vehicles to allow for a meaningful analysis of the distribution of error rates.
The results are shown in Figure \ref{fig:attacks-gini}; as before, each point is the mean of five runs, and the sample standard deviation is indicated.
Recall that in our setup, a Gini index closer zero means that the distribution of false positive/negative rates over the vehicles is closer to being equal, without making statements about the actual value.

\subsection{Discussion: Dispersion of Errors}
\label{sec:discussion}

In this assessment, the Gini index for false positives rates is the same for every attacker: we first discuss the false negative dispersion per attacker.
This dispersion gives us information about how different the detection performance is depending on the relative position of the attacker and the benign receiver.

For the ART, we observe that the dispersion of false negatives with regards to attacker 1 is very high: this can be explained by the fact that vehicles near the claimed constant false position will not be able to detect it with this mechanism.
A similar effect can be observed for attacker 4, while for the other attacks, the dispersion only increases when the threshold is very low.
This reflects the increasing recall discussed in the previous section, but remember that the precision also decreases significantly here.
The SSC has a very low dispersion of errors for attacker 2, but unfortunately this is also the attacker against which its' precision is very low.
Against attacker 8, the SSC outperforms the ART; the dispersion of errors suggests that this could be a localized effect, meaning that a combination of these mechanisms is likely to be feasible.
For the mechanisms that perform very poorly against certain attackers (SAW and DMV), the Gini index shows that their poor performance is not easily fixed: the error dispersion is very close to zero in most cases.
The exception is the DMV with regards to attackers 2 and 16: some performance improvement may be achievable by making changes to this detector.
Finally, note that for attacker 1, the recall of the DMV is very high, while the Gini index for the false negative rate is very close to zero.

One can observe that the dispersion of false positives for the ART show that for higher thresholds, the amount of false positives is significantly skewed over the population; this reflects the intuition that receiving a message from up to 700 meters away is unlikely but not impossible; however, for a threshold of 800, the dispersion is zero.
For the SSC, we observe that the threshold is much less relevant to the observed dispersion; this suggests that the mechanism would need to be changed more fundamentally to flatten the dispersion.
A notable case is the DMV: this mechanism has a very high Gini index, and thus is an excellent candidate for fusion with other sources.
In this particular setup, where detector assess the reliability of each message from the same source in isolation from other sources, an attack cannot lead to more false positives.
Another class of attacks, where an attacker aims to convince a benign vehicle of a false perception of the traffic scenario (e.g., claiming a traffic jam where there is none, by convincing the target that the average speed is much lower than it actually is), this is not necessarily the case.
Future work could use our metric to assess the real impact of this type of attack, as well as the use of this attack for \emph{data-driven bad mouthing attacks}: causing a benign vehicle to incorrectly classify another benign vehicle as malicious by convincing it of a false aggregate.

\section{Conclusion}
\label{sec:conclusion}

In this paper, we have introduced a new dataset for misbehavior detection in vehicular networks, called \emph{VeReMi}.
The purpose of this extensible, publicly available dataset is to provide a basis on which researchers can compare detection results in a wide set of traffic behaviors and attacker implementations.
We have additionally shown the application of this dataset to two existing, well-studied detection mechanisms (the acceptance range threshold and the sudden appearance warning), as well as two simple new detectors (simple speed check and distance moved verifier).
We have also provided a detailed discussion on why precision-recall is the preferred method of comparison, as well as a new metric that enables the user to determine which detectors can potentially be improved.
Using a combination of these metrics allows developers to have a more holistic view of a detector's assessment, which is information that can also be used in many fusion frameworks.
In our continuing work, we will use these metrics and this dataset as a basis to assess other commonly employed mechanisms, such as fusion between mechanisms and trust establishment.
This dataset will enable other researchers to compete with our detectors.

For future work, we see several directions beyond these detection performance improvements; one of these is to assess the feasibility of machine learning techniques for misbehavior detection.
The dataset can be used to either learn the attacker behavior (enabling high-quality detection of specific attacker patterns) or the benign behavior (enabling the detection of deviations from this behavior).
We expect that this direction is a feasible way to generate detector designs for specific scenarios that will also occur in the real world.
However, we caution against using this data (or even our simulation code) as the sole foundation for the evaluation of such machine-learned models.
Different real-world conditions (something as simple as a different speed limit on all roads) can impact the performance of such a learned model in a way that is not detectable without generating independent simulation results, or through the use of real-world data.

Future work should more closely investigate the available metrics from a security perspective.
Although PR graphs are considered advantageous over most other options, they clearly do not give a complete picture of detector performance.
The challenge in detection of malicious activity is that the difference between modeled behavior and observed behavior for both the attackers and the benign actors is fundamentally unknowable in advance of an attack.
For example, benign actors will likely transmit messages with significant GPS errors in areas where urban valleys exist, and attackers may develop new methods or tune their parameters to avoid detection.
Thus, although we feel that a dataset can function as a solid baseline for the behavior of different detection mechanisms, it is important to remark that such a dataset will always have the inherent limitation that the overall attacker prevalence is not generalizable.
This is what makes misbehavior and anomaly detection distinct from a simple classification task (such as medical diagnostics), for which the prevalence can be estimated -- dedicated metrics for misbehavior detection is likely the way forward.

\section*{Acknowledgement}
The authors thank Florian Diemer and Leo Hnatek for the contribution of several detector implementations in Maat, and Henning Kopp for discussions regarding the Gini index.
Experiments for this work were performed on the computational resource bwUniCluster funded by the Ministry of Science, Research and the Arts Baden-W\"{u}rttemberg and the Universities of the State of Baden-W\"{u}rttemberg, Germany, within the framework program bwHPC.
This work was supported in part by the Baden-W\"{u}rttemberg Stiftung gGmbH Stuttgart as part of the project IKT-05 AutoDetect of its IT security research programme.

\bibliographystyle{abbrv}
\bibliography{references}

%
%
%
%
%
%
%
%
%
%
\end{document}